\documentclass[final,5p,twocolumn,10pt,sort&compress]{elsarticle}

\usepackage[english]{babel}
\usepackage{enumerate}  
\usepackage[shortlabels]{enumitem} 
\usepackage{amsmath, amssymb, amsthm}
\usepackage{empheq} 
\usepackage{graphicx}
\usepackage[dvipsnames]{xcolor} 
\usepackage[colorlinks=true,linkcolor=NavyBlue,citecolor=NavyBlue,urlcolor=NavyBlue]{hyperref} 
\usepackage{appendix}
\usepackage{comment}
\usepackage{subcaption}
\usepackage{booktabs,dcolumn,caption}
\usepackage{tabularx}
\usepackage{multicol} 
\usepackage{balance}
\usepackage{xurl}
\usepackage{booktabs}   
\usepackage{siunitx}    

\usepackage[switch]{lineno}    

\makeatletter
\if@twocolumn
  \usepackage[switch]{lineno} 
\else
  \usepackage{lineno}         
\fi
\makeatother

\usepackage{rotating}   
\usepackage{adjustbox}  

\newcommand{\nt}[4]{\int \limits_{#1}^{#2}~\mathrm{d}#3~{#4}} 	

\makeatletter
\def\ps@pprintTitle{%
 \let\@oddhead\@empty 
 \let\@evenhead\@empty
 \def\@oddfoot{}%
 \let\@evenfoot\@oddfoot}
 \makeatother

\usepackage{titlesec}

\titleformat{\section}
  {\bfseries\Large} 
  {\thesection}{1em}{} 

\titleformat{\subsection}
  {\bfseries\normalsize} 
  {\thesubsection}{1em}{} 

\usepackage{xcolor}        
\newcommand{\coloredtext}[1]{{\color{black} #1}}

\begin{document}
\begin{frontmatter}

\title{\coloredtext{Social feedback amplifies emotional language in online video live chats}}

\author[1,2]{Yishan Luo}
\author[2]{Didier Sornette}
\author[2,3]{Sandro Claudio Lera\corref{cor1}}

\cortext[cor1]{corresponding author: \texttt{leras@sustech.edu.cn}}
\address[1]{Warwick Business School, University of Warwick, Coventry, UK}
\address[2]{Institute of Risk Analysis, Prediction and Management, Southern University of Science and Technology, Shenzhen, China}
\address[3]{Connection Science, Massachusetts Institute of Technology, Cambridge, USA}

\begin{abstract}
A growing share of human interactions now occurs online, where the expression and perception of emotions are often amplified and distorted.  
Yet, the interplay between different emotions and the extent to which they are driven by external stimuli or social feedback remains poorly understood.  
We calibrate a multivariate Hawkes self-exciting point process to model the temporal expression of six basic emotions in YouTube Live chats.  
This framework captures both temporal and cross-emotional dependencies while allowing us to disentangle the influence of video content (exogenous) from peer interactions (endogenous).  
We find that emotional expressions are up to four times more strongly driven by peer interaction than by video content.  
Positivity is more contagious, spreading three times more readily, whereas negativity is more memorable, lingering nearly twice as long.
Moreover, we observe asymmetric cross-excitation, with negative emotions frequently triggering positive ones, a pattern consistent with trolling dynamics, but not the reverse.  
These findings highlight the central role of social interaction in shaping emotional dynamics online and the risks of emotional manipulation as human-chatbot interactions become increasingly realistic.
\end{abstract}


\end{frontmatter}

\section*{Introduction}
\label{sec:intro}

In recent decades, a growing share of human interaction has shifted to online platforms.  
The distinctive features of these environments, such as anonymity, reduced accountability, and the absence of non-verbal cues, shape digital interactions in ways that differ markedly from face-to-face communication \cite{goldenberg2023amplification}.  
This transformation has spurred substantial interest in understanding the role of emotions in online engagement and their broader societal consequences.  

Emotional contagion in digital settings is well documented:
Both positive and negative emotions propagate through social networks \cite{kramer2014, ferrara2015, coviello2014, gruzd2011, he2016}, 
and emotional content increases the likelihood of resharing \cite{chen2022, ferrara2015}.  
Moral emotions, in particular, enhance the spread of moral and political discourse \cite{brady2017}, and the strength of social ties modulates emotional diffusion: 
strong ties amplify influence \cite{fan2018}, while anger spreads more effectively than joy through weak ties, reaching wider audiences \cite{fan2020}.  

Platform design also shapes emotional dynamics. 
Digital media platforms are motivated to upregulate user emotions \cite{goldenberg2020digital}, 
and large-scale data from YouTube suggests that live-streaming environments can further intensify emotions through mechanisms such as shared attention \cite{luo2020emotional}.  
Temporal studies show that positive emotions tend to rise quickly and fade fast, whereas negative emotions build more gradually and persist longer \cite{naskar2020, fan2019, pellert2020}.

Emotional dynamics on social platforms also affect collective behavior and discourse by amplifying polarization, incivility, and misinformation \cite{graciyal2021, buder2021, castano2021internet, coe2014online, prollochs2021, avalle2024persistent}.  
Negative emotions are particularly predictive of fake news engagement during crises, such as the COVID-19 pandemic \cite{farhoudinia2024emotions}.  
Emotional expression in online reviews is also a good predictor of commercial outcomes across domains including movie and restaurant success \cite{rocklage2021mass}.

Despite growing evidence of emotional contagion in digital spaces, we identify four key limitations in the literature.
First, previous studies have found evidence for both positive and negative biases on online platforms, highlighting the need for further investigation to compare, contextualize, and reconcile these different effects \cite{goldenberg2020digital, paletz2023emotional}. 
Online environments foster a positivity bias driven by self-representation and social validation mechanisms \cite{schreurs2021introducing, sheldon2016instagram, spottswood2016positivity}.
Meanwhile, negativity can spread more easily due to cognitive biases that prioritize negative information \cite{rozin2001negativity, knobloch2020confirmation, soroka2019cross}.
Previous studies have found that positivity tends to be more contagious on online platforms \cite{coviello2014, ferrara2015measuring, gruzd2011happiness}.
Meanwhile, recent findings suggest that while negativity often has greater reach \cite{robertson2023negativity, schone2023negative}, emotional virality is complex and context-dependent \cite{paletz2023emotional}.
These mixed results call for a reconciliation of our understanding of the different ways in which negative and positive emotions spread within social networks.

Second, while emotional contagion is well documented, less is known about how different emotions trigger one another in large-scale social interactions. 
Previous work has focused on individual emotion transitions \cite{kuppens2010feelings, pe2012dynamic} and interpersonal dynamics, where emotional expressions can elicit both mimicry and divergent responses depending on the context \cite{hareli2008emotion, thornton2017mental}. 
In addition, previous studies have used co-occurrence patterns of emotions to improve emotion classification using natural language processing (NLP) models \cite{wang2020, chou2022}.
A graph-based approach incorporating emotion correlations has been found to outperform previous benchmarks in emotion classification tasks \cite{xu2020emograph}. 
However, how different emotions trigger each other in the context of online social interactions is relatively underexplored.

Third, with few exceptions \cite{naskar2020, fan2019, pellert2020}, most studies do not account for the subtle temporal dependencies underlying emotional contagion.  
Contagion is often measured in terms of whether content is shared or reciprocated, while the timing between emotional expressions receives comparatively little attention.  

Fourth, the mechanisms by which online engagement shapes users' emotional states remain inadequately understood.
Although prior studies have shown that emotions are widely spread through social networks, it is still unclear whether this contagion is primarily content-driven arising from the emotional tone of the information, or peer-driven, shaped by social reinforcement and interactions among users.
For example, live comments during real-time events exhibit greater emotional intensity compared to standard comments \cite{luo2020emotional}, but the extent to which this amplification stems from the emotional nature of the event versus social interactions among participants is unclear.
More broadly, distinguishing between exogenous (e.g., content-driven) and endogenous (e.g., peer-driven) influences in online emotion contagion remains a fundamental challenge in the study of complex social systems \cite{Sornetteendoexo05}.

\begin{figure*}[!htb]
	\centering
	\includegraphics[width=0.95\textwidth]{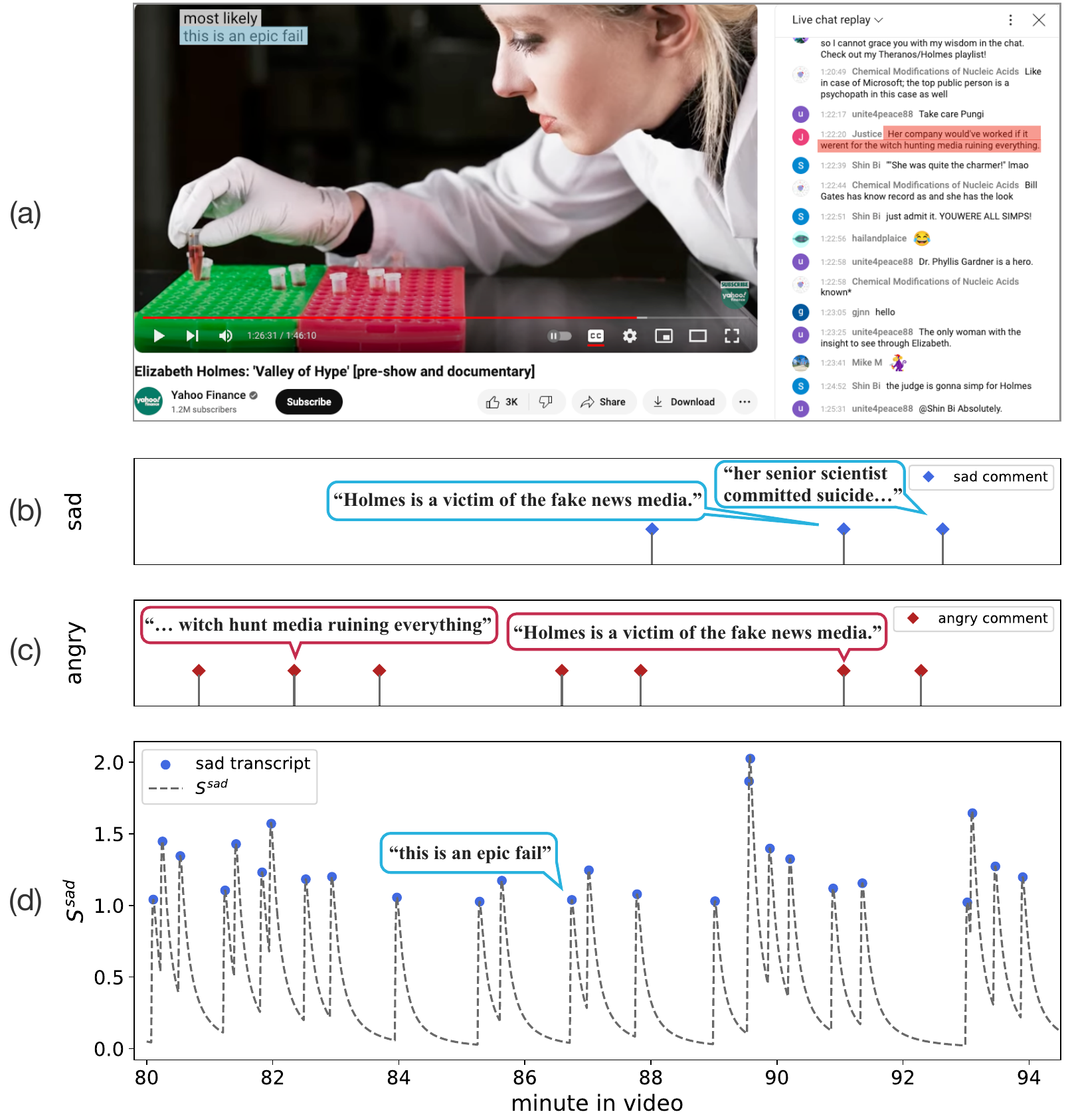}
	\caption{
    	(a) 
    	Screenshot of a YouTube live video on the \textit{Theranos scandal} involving Elizabeth Holmes with a live chat section on the right.
    	The video content is captured by transcripts (\textit{subtitles}), which we use to proxy exogenous emotional stimuli for the live chats.
        The live chat section displays timestamped messages from users reacting to the video content in real-time as the live video streams.
        We highlight an example of a transcript labeled as \textit{sad} in blue, and a live chat message labeled as \textit{angry} in red.
        (b,c)
        We visualize the extraction of emotions from the live chat in the screenshot above.
        We plot a subset of live chat messages from the video sample that are labeled as \textit{sad} (\textit{angry}), indicated with diamond markers in blue (red).
        The shared x-axis shows the time in the video in units of minutes.
        The y-axis is unit-less. 
        We label emotions non-exclusively.
        For instance, the sentence ``Holmes is a victim of the fake news media.'' is labeled as both sad and angry. 
        We assume these emotions are generated by the latent, inhomogeneous intensity defined by expression \eqref{eq:multi_hawkes_general_main}. 
        (d)
        Time-varying component of the emotion \textit{sad} in the video (transcript). 
        Blue dots annotate the arrival of transcripts that are labeled as \textit{sad}, shown with a 2-second rightward shift for alignment with peaks in the signal for visual convenience (see Methods section for details).
        The black dashed line represents the temporal function $S^\text{sad}(t)$, capturing the presence of sad emotions within the video.}
	\label{fig:livechat}
\end{figure*}

We address these gaps by modeling emotional contagion as a sequence of discrete events—moments in time when users express distinct emotions in response to stimuli or peer interaction.  
Such event-based dynamics are naturally captured by self-exciting point processes, which allow past events to influence the probability of future ones.  
This temporal dependency and recursive structure make the multivariate Hawkes process a particularly suitable tool, as it explicitly accounts for feedback effects between events over time - a feature that standard regression approaches typically do not accommodate \cite{saichev2011generating, saichev2013hierarchy}. 

Using this framework, we analyze the joint dynamics of six basic emotions in YouTube live chats. 
We define the system of interest as the collective set of viewers participating in the live chat. 
Under this group-level framework, emotional expressions triggered by prior messages within the chat are considered endogenous, while influences originating outside the chat, primarily from the video content, are treated as exogenous. 
This distinction allows us to study how social interaction shapes group-level emotion dynamics during livestreams. 

Concretely, we classify emotions originating from video content as exogenous (\textit{exo}) inputs and those expressed in the chat as endogenous (\textit{endo}) responses driven by peer interaction.  
\coloredtext{
Building on prior evidence of emotional contagion in digital environments, this study examines how emotional dynamics in live discussions arise from the interplay between external video content and social feedback.  
We address the extent to which emotional expressions are driven by exogenous video stimuli versus endogenous peer interactions.
We further quantify the self- and cross-contagion effects as well as the temporal interdependencies of six different emotion types. 
These aims guide our empirical modeling of emotion dynamics using a multivariate Hawkes process framework applied to large-scale YouTube live chat data.
}

\section*{Methods}
\label{sec:methods}

\coloredtext{
There was no preregistration for any of the studies in this work.
}


\subsection*{YouTube Live Chat Data}
\label{sec:data_collection}

We examine the dynamics of emotions in online discussions through a collection of \textit{YouTube Live} videos.
YouTube is a video-sharing platform that caters to a wide range of users across different cultures.
YouTube Live allows the audience to chat with each other while the video is being played in real time.
We refer to live discussions as \textit{live chats} and a single user live comment as a \textit{live chat message}, or just a message. 
Videos with the \textit{live chat replay function} record time-stamped live discussions and allow viewers to retrieve the live experience by simultaneously replaying the video content and live chat messages.
Figure \ref{fig:livechat}(a) shows an example of a YouTube live video (left) with a live chat section (right) where viewers post messages visible to the entire audience as the video streams.

We examine emotion dynamics in YouTube live videos with replay functionality across 27 topics ranging from comedy and documentary to sports and travel (see Supplementary Note 1 for the complete list).
We first compile the list of topic keywords and collect completed live video IDs for each topic keyword using the keyword search function via the YouTube API in Python.
We initially extract both the video transcripts and live chat messages associated with each video. 
This process yields two distinct lists: one of time-stamped chat messages and another of time-stamped video transcripts (\textit{subtitles}). 
Notably, YouTube Live Chat offers two display modes: an unfiltered chronological stream (Live Chat) and a filtered version (Top Chat) that removes spam and inappropriate messages.
We use the Live Chat version in our data collection process, ensuring that all messages are collected without algorithmic filtering. 
While viewers can choose different display modes when watching the video, we note that YouTube's filtering is not personalized and applies uniformly across all viewers.
This stands in contrast to the personalized recommendation algorithms commonly used on other social media platforms, which can introduce more substantial distortions in the content users are exposed to.

We proceed by selecting only those videos in which the median interval between successive live chat messages ranges from 1 second to 5 minutes. 
Additionally, we disregard videos where more than 70\% of the chat messages are in a language other than English. 
For the videos retained, we eliminate non-English live chat messages. 
Furthermore, we delete any chat messages that precede the first or succeed the last transcript time in each video. 
Ultimately, this results in a collection of 673,551 live chat messages from 1,957 videos.

\subsection*{Labeling Emotions in Text} 
\label{sec:emotion_labels}

We consider expressions of emotions through text.
Text-based emotion measurement has been shown to correlate with self-reported emotions, establishing it as a reliable and valuable method for capturing emotional expressions \cite{munin2025language, tov2013detecting}.
We delineate emotions according to the commonly used \textit{ 6 basic emotions model} \cite{ekman1992}.
These fundamental emotions are 
$
\mathcal{E} \equiv \{\text{joy, surprise, anger, disgust, fear, sadness}\}
$. 
The discrete emotion model remains a favored method within affective computing research \cite{calvo2010affect}. 
Notably, the classification of emotions is effective for emotion analysis in textual content \cite{calvo2013emotions}. 
Moreover, research indicates that discrete emotion modeling often surpasses dimensional models in elucidating the sharing dynamics of online content, 
thereby more accurately capturing users' emotional displays in digital settings \cite{paletz2023emotional}.

Multiple emotions can arise simultaneously in a single sentence \cite{wang2020, xu2020emograph}.
We thus assign to each text a vector of length 6 indicating the non-exclusive presence of each of the six emotions.
This amounts to a multilabel classification problem. 
For training data, we rely on the SemEval-2018 dataset containing 6,838 Tweets with non-exclusive emotion labels across 11 emotions \cite{mohammad2018}.
We group semantically related emotions into six core categories by mapping anticipation, optimism, love, and trust to joy; pessimism to sadness; and retaining sadness, anger, disgust, fear, and surprise in their original categories.
Subsequently, we fine-tune a \textit{Roberta} transformer model \cite{liu2019roberta}.
The transformer takes as input a text and gives as output the probabilities for each of the 6 basic emotions via 6 independent sigmoid activations. 
We retain emojis in the text input, as they serve as an important channel for emotional expression in online communication.
We use the binary cross-entropy loss function for training.
During prediction, we convert each sigmoid output to a binary indicator at the 0.5 cutoff.
This allows us to assign to each live chat message and video subscript a subset of the emotions $\mathcal{E}$.
We provide examples of live chat messages and the corresponding assigned emotion labels in Supplementary Note 1.

For every video and emotion analyzed, we determine the average count of live chat messages per minute. 
To exclude outliers or inconsistencies in live chat data, only those videos where this average falls between the $20^{\text{th}}$- and $80^{\text{th}}$-quantiles of all videos are included. 
This process results in a compiled set of 92,412 live chat messages that span 397 videos, totaling 780 hours. 
The median video length is 107 minutes and per video, the median number of live chat messages is 204. 
When considering all videos, the median number of live chat messages per minute is as follows:
1.63 for joy, 0.17 for disgust, 0.16 for anger, 0.11 for sadness, and for both fear and surprise, 0.03. 
Additional details are provided in Supplementary Note 1.

\subsection*{Emotions as Self-Excited Point Processes} 
\label{sec:hawkes}

The treatment of emotions as discrete events makes point processes a natural analytical framework. 
A point process models the occurrence of discrete events in continuous time, with the simplest case being the Poisson process, where events occur independently at a constant rate.  
However, emotions are not independent; they exhibit self-excitation, which means that past emotions increase the likelihood of future ones \cite{prochazkova2017connecting}.
To account for this, we adopt the \textit{Hawkes process}, one of the simplest self-exciting point processes, which extends the Poisson process by incorporating memory and self-excitation while maintaining a linear and additive structure \cite{hawkes1971spectra}.  
The intensity function of a Hawkes process is given by
$
\lambda(t) = \mu + \sum_{t_i < t} g(t - t_i)
$
where
$\mu$ is the baseline event rate, and $g(t - t_i)$ is a triggering function that quantifies the influence of past events that occurred at discrete times $\{ t_j \}$ on future events.  
This formulation allows for a \textit{branching process} interpretation, where events can be classified into exogenous (generated by external stimuli) and endogenous (triggered by past events) \cite{hawkes1974cluster}.    
Hawkes processes have been widely applied to self-exciting phenomena, including 
earthquakes \cite{nandan2021}, 
financial transactions \cite{blanc2017quadratic, wehrli2021},
video viewing activities \cite{crane2008robust, rizoiu2017expecting},  
and information cascades on social networks \cite{kobayashi2016tideh, kong2021evently}.  

Emotional contagion, the phenomenon in which emotions spread from one individual to another, is naturally captured by the self-exciting nature of the Hawkes process \cite{prochazkova2017connecting}.
In our context, emotions are modeled as \textit{events} whose statistical occurrence follows a point process.
Because we have more than just one emotion and the emotions (presumably) cross-excite each other, we represent the dynamics of $|\mathcal{E}|=6$ distinct emotions using a multivariate Hawkes process \cite{laub2021elements}.  
It is our goal to understand to what extent a given emotion $f \in \mathcal{E}$ triggers another emotion $e \in \mathcal{E}$ either exogenously from the video to the live chat or endogenously from live chats to live chats. 
Here, the adjectives \textit{exogenous} and \textit{endogenous} refer to the point of view of live chat:
If emotion $f$ appears in the video and triggers an emotion $e$ in the live chat, we call it exogenous (\textit{exo}). 
In contrast, if emotion $f$ also appears in the live chat and subsequently triggers emotion $e$ in the live chat, we call it endogenous (\textit{endo}).
We assume that emotions in the video and in the live chat can trigger emotions in the live chat, 
but that emotions in the live chat cannot influence emotions in the video. 
We demonstrate in Supplementary Note 5 that our results remain qualitatively consistent both with and without the subset of videos for which this assumption does not necessarily hold.

In a more formal representation of these concepts, we label $\{ t_j^e \}$ as the set of timestamps corresponding to when a live chat message expressing emotion $e$ was sent.
We assume that this set of event times is sampled from a latent inhomogeneous event intensity $\lambda^e(t)$. 
The intensity is determined by three factors: 
(i) a uniform, constant exo base rate $\mu^e_0$, reflecting the spontaneous generation of live chat messages of emotion $e$ without any prompting from the video or chat; 
(ii) a non-uniform, dynamic exo rate $\mu^{e}_1(t)$, representing the generation of live chat messages of emotion $e$ triggered by events in the video; 
(iii) an endogenous rate considers that earlier live chat messages provoke subsequent ones. 
This is defined by the following expression for the intensity of emotion $e \in \mathcal{E}$ as a Hawkes point process:
\begin{equation}
\scriptstyle 
	\lambda^{e}(t)
	= 
	\overbrace{
	\mu^e_0
	}^{\text{exo base rate}}
	+ 
	\overbrace{
	\sum_{f \in \mathcal{E}} \nu^{e, f} S^f(t)
	}^{\text{exo video influence}~\mu_1^e(t)}
	+ 
	\overbrace{
	\sum_{f \in \mathcal{E}} \sum_{t_j^{f} < t} \phi^{e, f} \left( t-t_j^{f} \right).
	}^{\text{endo chat influence}}
	\label{eq:multi_hawkes_general_main}
\end{equation} 
Intensity is defined such that $\lambda^{e}(t) \mathrm{d}t$ is the probability that an emotion $e$ occurs in the live chat between $t$ and $t+\mathrm{d}t$.
Equation \eqref{eq:multi_hawkes_general_main} states that the intensity at which emotions of type $e$ are generated in the live chat is an additive function of past emotions. 
The term $\mu_0^e$ is just a constant as described above. 
The term $S^f(t)$ is the time-varying presence of emotion $f$ in the video subscript (see next section for details and Figure \ref{fig:livechat}(d) for an example of $S^\text{sad}(t)$).
The term $\nu^{e,f} S^f(t)$ then represents the rate at which emotion $f$ in the video triggers emotion $e$ in the live chat. 
For example, a sad scene in the video can trigger an angry message in the live chat.
Summing over all emotions $\mathcal{E}$ we arrive at the rate $\mu_1^e(t)$. 
This rate $\mu_1^e(t)$ captures the rate at which any emotion in the video triggers emotion $e$ in the live chat. 
The third term captures the endogenous influence of previous live chat messages.
Here, $\phi^{e, f}(t-t_j^{f})$ represents the influence of a past emotion $f$ at time $t^f_j$ on the likelihood of observing the emotion $e$ at time $t$. 
The intensity kernel $\phi^{e,f}(\cdot)$ is monotonically decaying, so the more time has passed, the less influence a prior message has.
We follow the common assumption that $\phi$ is exponential,
$\phi^{e, f}(t) = \left. \alpha^{e, f} e^{- t / \gamma^e} \right/ \gamma^e$, 
with decay time $\gamma^e$.
The larger $\gamma^e$, the longer the direct memory of a previous live chat message has on emotion $e$. 
Here, we have assumed $\gamma^{e,f} \equiv \gamma^e$ for simplicity, but generalizations are straightforward, albeit at the cost of increasing the number of parameters to be estimated. 
The coefficient $\alpha^{e,f}$ represents the endogenous excitation effect of emotion $f$ on emotion $e$ in the live chat. 

In summary, the intensity $\lambda^e(t)$ from Equation \eqref{eq:multi_hawkes_general_main} is parametrized by a set of $2 |\mathcal{E}| + 2=14$ parameters: 
the exogeneous rate $\mu^e_0$, the decay time $\gamma^e$, as well as $|\mathcal{E}|$ parameters for $\nu^{e,f}$ and $\alpha^{e,f}$, respectively.
We estimate these parameters by maximizing their log-likelihood functions (see below).
Since each emotion has its own intensity $\lambda^e$, our model has a total of $|\mathcal{E}| \left( 2 |\mathcal{E}| + 2 \right)=84$ parameters. 
However, the estimations of the different $\lambda^e$ are decoupled from each other, so that we can estimate each set of $12$ parameters separately. 
The fitted parameters are visualized in Figure \ref{fig:result} and discussed in the Results section. 

We finally note that, in our analysis, we treat the aggregate of all viewers in the live chat as the endogenous system. 
At this level of description, any emotional comment triggered by prior comments in the chat is modeled as endogenous, while the influence of video content and a small baseline rate are treated as exogenous. 
This group-level framing is appropriate given our empirical setting, where repeated messages from the same viewer occur much more sparsely than messages exchanged across different viewers. 
If sufficient data were available, one could estimate a multivariate Hawkes process with \(6N\) dimensions, where each of the \(N\) viewers can experience six basic emotions. 
This would allow a precise separation between intra-individual self-excitation (diagonal terms of the kernel) and inter-individual contagion (off-diagonal terms). 
However, in such a specification, the number of parameters scale like $N^2$, which is computationally intensive and not feasible given our current dataset, where the average viewer posts only about nine messages per video. 
An intermediate approach would be to assume homogeneity across users and fit a single six-dimensional Hawkes process to all data, pooling observations across individuals. 
This would make it possible to quantify the average influence of other viewers and the video on a given individual, while keeping the parameter space manageable. 
Exploring this alternative model formulation represents a promising avenue for future research.

\subsection*{Parametrization of Video Influence} 
\label{sec:video_derivation}

We model the arrival of emotions as a multivariate Hawkes process given by expression \eqref{eq:multi_hawkes_general_main}.
For emotion $e \in \mathcal{E}$, within a given live chat session, we observe $N^e$ events, with the $j^{\text{th}}$ event taking place at time $t_j^{e}$.
We define $t_0^e \equiv 0$, $t_{N^{e}+1}^e \equiv T$ where $T$ is the duration of the video.
In this way, the observation period for the entire process is $\{ t_i^e \mid i = 0, \ldots, N^{e}+1 \}$.
Similarly, we denote by $\tau^e_i$ the time at which the $i$-th subscript of emotion $e$ appears in the video. 
We express time in units of minutes throughout the analysis and event times are treated as continuous variables throughout the model fitting process. 
The live chat messages have a temporal resolution measured in milliseconds while the video transcripts are measured in centiseconds.
For example, an event occurring at 5 minutes and 3.2 seconds is recorded at 5.0533. 

In the remainder of this subsection, we describe the process of parameterizing the time-varying exogenous video influence $S^f(t)$. 
In essence, $S^f(t)$ denotes the level of emotion $f$ exhibited in the video at a particular time $t$ (refer to Figure \ref{fig:livechat}(d) for an illustration). 
Readers who are mainly interested in the qualitative results of our study may skip this and the next subsection and continue with the Results Section. 

Previous studies have shown that assuming the exogenous influence $\mu^e(t)$ to be constant can lead to false attribution to endogenous effects \cite{wheatley2019, wehrli2021} (see Supplementary Note 2 for additional confirmation).
The YouTube live videos provide us with the unique opportunity to observe a time-varying influence of video content on user discussions.
We use the video transcript to capture the time-varying emotional content of the video.
For example, it is reasonable to assume that a particularly sad scene in the video induces sad emotions in the live chat.
In addition, cross-influence can be expected; for example, a sad scene in the video induces angry emotions in the live chat (Figure \ref{fig:livechat}).
In particular, this is to be distinguished from a sad scene in the video triggering a sad message in the live chat, which in turn triggers either a sad or angry message. 
To capture such cross-influence from the videos, we parameterize the exogenous intensity of emotion $e$ as
\begin{equation}
    \mu^e(t) = \mu^e_0 + \mu^e_1(t) \equiv \mu^e_0 +  \sum_{f \in \mathcal{E}}~ \nu^{e, f} ~S^f(t),
    \label{eq:mue(t)}
\end{equation}
where $\mu^e_0$ is a time-invariant, unobserved baseline intensity of spontaneous expression of emotion $e$ in the absence of influence from video or prior live chat messages.
By contrast, we denote by $S^f(t)$ the time-varying intensity of emotion $f$ in the video,
and $\nu^{e,f}$ the cross-influence from emotion $f$ in the video to emotion $e$ in the live chat. 
The term $\nu^{e,f} ~S^f(t)$ is thus the time-varying intensity at which emotions of type $f$ in the video triggers emotions of type $e$ in the live chat.

The intensity $S^f(t)$ is obtained as an interpolation of observed emotions $f$ in the subtitles of the video. 
More formally, 
$S^f(t) = \sum_{\tau^f_j < t} s^f_{\tau_j}(t)$ 
where $\left\{ \tau^f_j \right\}$ enumerates all the times at which a subtitle of emotion $f$ appears in the video.
Here, $s^f_{\tau_j}(t)$ captures the time-varying influence of a video subtitle of emotion $f$ appearing in the video at time $\tau_j$. 
Clearly, $s^f_{\tau_j}(t) = 0$ for $t < \tau_j$. 
In order to incorporate the video impact, we need to make reasonable assumptions on the shape of
the temporal dependence of the influence of emotions in the video on emotions in the live chat.
It is natural to assume that it takes a few seconds for the audience to process the transcript and react. 
We also expect the influence of the video to peak after a rapid increase in intensity and then fade out with time.
Empirical evidence demonstrates that information retention in humans decays with a fat tail, that is,
slower than an exponential \cite{pollmann1998forgetting}.
To approximately capture the rapid initial increase and subsequent slow decline, 
we use the log-normal function to parametrize the memory of a subscript in the video:
\begingroup
\small
\begin{equation} 
    s^f_{\tau_j} = \frac{1}{\sqrt{2\pi}\sigma (t - \tau_j)} \exp\left(-\frac{(\ln (t - \tau_j) - \mu)^2}{2\sigma^2}\right).
\label{eq:s_parametrization}
\end{equation} 
\endgroup
Under a finite range of the variable that depends on $\sigma$, the log-normal function exhibits shapes similar to a power law \cite{sornette2006critical}, 
while also displaying an initial steep increase. 

To fix $\mu$ and $\sigma$, we assume that
the intensity of each transcript peaks $2$ seconds after appearance and that
50\% of the emotion intensity for each transcript is manifested within $10$ seconds of transcript appearance.
In other words, we assume that the maximum of the log-normal function lies at $2$ seconds, $\exp(\mu - \sigma^2) = 2$, and that the median is equal to $10$ seconds, $\exp(\mu) = 10$.
Numerically solving for $\mu$ and $\sigma$ yields $2.3$ and $1.3$ respectively in units of minutes.
To ensure that our results are not strongly dependent on this choice of $\mu$ and $\sigma$, 
we have checked that our results remain qualitatively similar for different values and functional shapes (see Supplementary Note 5). 
Alternatively, these parameters could be directly fitted from the data, a task which we leave for future research. 

In summary, the shape $S^f(t)$ is a sum of log-normal functions, with local peaks at $2$ seconds after the appearance of a new subtitle of emotion $f$.
An example of $S^{\text{sad}}$ is shown in Figure \ref{fig:livechat}(d). 
We further stress that our framework is different from the typical multivariate Hawkes process in that expression \eqref{eq:mue(t)} contains cross-influence terms from different exogenous sources.
This is because we condense a bi-multivariate Hawkes model containing two systems of events (video and chat) into one multivariate Hawkes self-excited conditional point process for the chat. 
This specification is therefore necessary to distinguish video-based emotion events and chat-based emotion events.

\subsection*{Fitting the Parameters of the Point Process} 
\label{sec:log-likelihood}
A general representation of the log-likelihood of intensity $\lambda^e$ from \eqref{eq:multi_hawkes_general_main} is given by
\begin{subequations}
\begin{align}
	\log \left( L^{e} \right)
        &= \sum_{i=1}^{N^{e}} \log(\lambda^{e}(t_i^e)) -\nt{0}{T}{s}{\lambda^{e}(s)} \\
        &= \sum_{i=1}^{N^{e}}\log(\lambda^{e}(t_i^e)) - \nt{0}{T}{s}{\mu^{e}(s)} -  \nonumber \\
        & \quad \sum_{f \in \mathcal{E}} \sum_{t_j^{f}}\nt{t_j^{f}}{T}{s}{\phi^{e, f}(s)}.
        \label{eq:multi_llh_general_main}
\end{align}
\end{subequations}
We follow the common assumption that $\phi$ is exponentially decaying and write 
$\phi^{e, f}(t) = \left. \alpha^{e, f} e^{- t / \gamma^e} \right/ \gamma^e$
with decay time $\gamma^e$, such that \eqref{eq:multi_llh_general_main} can be expanded into
\begingroup
\small
\begin{equation}
\begin{aligned}
	&\log L^{e} \left( \mu^{e}_0, \nu^{e, f}, \alpha^{e, f}, \gamma^e \right) 
        = \\ 
        &\sum_{i=1}^{N^{e}}\log \left( \mu^{e}_0 + \sum_{f \in \mathcal{E}} \nu^{e, f}S^{f}(t_i) + 
        \sum_{f \in \mathcal{E}} \sum_{t_j^{f} < t} \alpha^{e, f} \frac{1}{\gamma^e} e^{-\frac{1}{\gamma^e} (t_i - t_j^{f})} \right) \\
        & - \mu^{e}_0 T - \sum_{f \in \mathcal{E}}\nu^{e, f}M^f_1 + \sum_{f \in \mathcal{E}} \sum_{t_j^{f}}\alpha^{e, f} \left(e^{-\frac{1}{\gamma^e} (T - t_j^{f})} - 1 \right),
        \label{eq:multi_cross_mu_main}
\end{aligned}
\end{equation}
\endgroup
where $M^f_1 \equiv \nt{0}{T}{s}{S^{f}(s)}$ is a constant that we calculate numerically.
A detailed derivation of Equation \eqref{eq:multi_cross_mu_main} is found in Supplementary Note 3.

We estimate the values for $\mu^{e}_0$, $\nu^{e, f}$, $\alpha^{e, f}$, and $\gamma^e$ maximizing Equation \eqref{eq:multi_cross_mu_main} using Quasi-Newton optimization with constraints. 
The variable bound constraints are specified as follows:
\(0 \leq \alpha \leq 50\), \(10^{-6}  \leq \nu \leq 10\), \(0.1 \leq \gamma \leq 20\), \(0 \leq \mu \leq 50\).

\begin{figure*}[!htb]
	\centering
	\includegraphics[width=\textwidth]{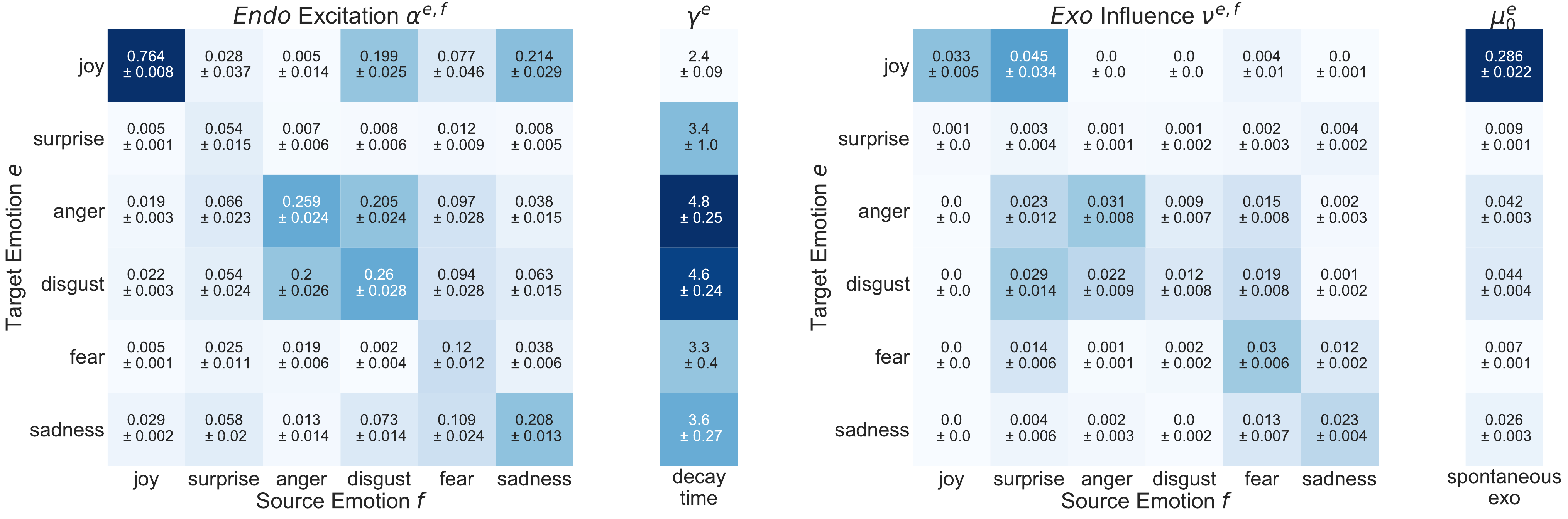}
	\caption{Visualizations of estimated parameters $\nu^{e,f}$, $\mu^{e}_0$, $\alpha^{e,f}$, and $\gamma^e$ of the intensity 
	defined by expression \eqref{eq:multi_hawkes_general_main} from the maximization of the log-likelihood function \eqref{eq:multi_cross_mu_main}.
             The parameters are fitted ten times, each time sampling 60\% (238) of the total 397 videos at random. 
             Values and error bars are then obtained as sample mean and standard deviations across all 100 fits, respectively.                 
             Entry $\alpha^{e,f}$ of the $\alpha$ matrix represents excitation from emotion $f$ to emotion $e$ in the live chat, 
             while $\gamma^e$ represents the characteristic time-scale over which past emotions trigger new emotions $e$. 
             The $\nu$ matrix demonstrates how emotions in the video trigger emotions in the live chat, 
             while $\mu_0$ illustrates the spontaneous baseline intensity of spontaneous emotion expression. 
             Note that the $\gamma^e$ and $\mu_0^e$ columns share the same $y$-axis emotion labels as the $\alpha$ and $\nu$ matrices, respectively.
		}
	\label{fig:result}
\end{figure*}

Note that Equation \eqref{eq:multi_cross_mu_main} represents the log-likelihood for a single video, $\log L^e = \log L^e_k$ where $k$ enumerates the videos in our data sample. 
Thus, we aggregate $\log L^e_k$ across all videos and estimate the model parameters by maximizing $\sum_{k} \log L^e_k$.
The total number of parameters is $2 |\mathcal{E}| + 2 = 14$ per emotion. 
This amounts to a total of $|\mathcal{E}| \left( 2 |\mathcal{E}| + 2 \right) = 84$ parameters to estimate for $|\mathcal{E}|=6$ emotions. 
However, the parameters for each emotion can be estimated independently because there is no interdependence between the parameters between different types of emotions.
This gives rise to six independent fits of $14$ parameters each. 
To verify that the obtained fits are valid, we first test our model with synthetically generated data and obtain reliable performance (see Supplementary Note 4).
We also refer to Supplementary Note 5 for additional robustness checks with respect to different parameterizations and subsets of emotions. 
To mitigate potential biases, we estimate our model parameters using bootstrapped samples.
We bootstrap our data $100$ times, each time sampling only 60\% of all available videos. 
This allows us to estimate the parameters as averages across bootstrapped samples, and we use standard deviations as error bars. 
The fitted parameters are shown in Figure \ref{fig:result}.
These parameters are interpreted in the Results Section and discussed more qualitatively in the Discussion Section.

\section*{Results} 
\label{sec:results}

\subsection*{Quantifying Exogenous and Endogenous Influences}
\label{sec:endo-exo}

\begin{figure*}[!htb]
	\centering
	\includegraphics[width=\textwidth]{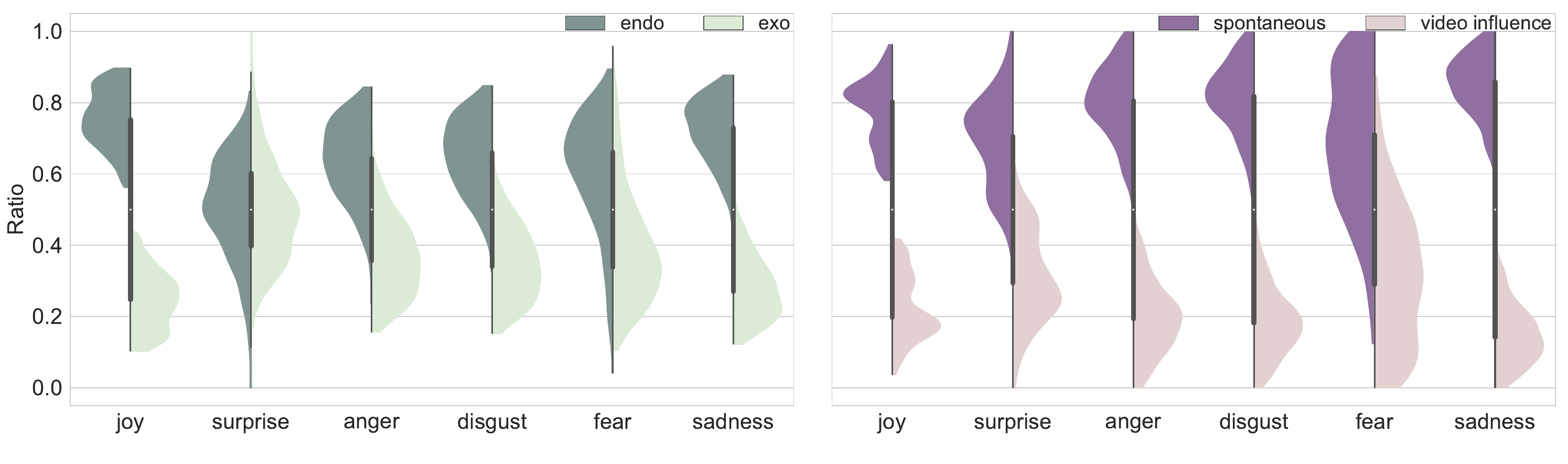}
	\caption{  (a)                
                For each video, we calculate the average ratio of endogenous (exogenous) intensity to the total intensity across time. 
                Plot (a) shows the distribution of these ratios across \coloredtext{$n = 397$} videos. 
                We notice that emotions are predominantly triggered endogenously. 
                (b)
                Same as plot (a) but for the ratio of spontaneous (video-influenced) intensity to total exogenous intensity.
		      }
	\label{fig:ratios}
\end{figure*}

Videos provide a shared experience among viewers, resulting in increased emotional intensities during live broadcast events \cite{luo2020emotional}.
However, it is unclear to what extent the emotional experience and amplification stem from the live videos themselves or the interactions among peers \cite{luo2020emotional}.
The Hawkes framework \eqref{eq:multi_hawkes_general_main} allows us to answer this question by comparing exogenously and endogenously triggered events.
The exogenous component, $\mu^{e}(t) = \mu^e_0 + \mu_1^e(t)$, represents influences from video content and elsewhere.
The endogenous component, $\sum_{f \in \mathcal{E}} \sum_{t_j^{f} < t} \phi^{e, f} \left( t-t_j^{f} \right)$, captures the influence of previous live chat messages and is therefore derived from interactions with peers. 
At any given time, we can determine the fraction of exo-influence as 
$
R^e_{\text{exo}}(t) =\left . \mu^e(t) \right/ \lambda^{e}(t)
$
and the ratio of endo influence as 
$
R^e_{\text{endo}}(t) = 1- R^e_{\text{exo}}(t) 
$
where all quantities are calculated with the estimated parameters (shown in Figure \ref{fig:result} and further discussed below).
Subsequently, we calculate the average exo- and endo-ratios, 
$\langle R^e_{\text{exo}} \rangle$ 
and 
$\langle R^e_{\text{endo}} \rangle$, 
of a given video by averaging over time.
To put it differently, for each video, $\langle R^e_{\text{exo (exo)}} \rangle$ represents the mean portion of intensity for emotion $e$ that is elicited exogenously (endogenously).
Figure \ref{fig:ratios} (left) shows the distribution of these quantities over all videos in our sample.
We observe that despite the constant exogenous influence from video feeds, endogenous influence dominates the emotion dynamics across emotion types.
This shows that emotional expressions in live discussions are disproportionately driven by social interactions with active participants rather than passive consumption of video content.
In particular, we see that the emotion of joy has the highest ratio of endogenous influence, being on average 75\% driven by prior participants. 
In other words, joy is three times more likely to be triggered endogenously than exogenously. 
In contrast, and according to common sense, the emotion of surprise is the most responsive to exogenous stimuli, where roughly 50\% of the total intensity is triggered exogenously.

Similarly, we can calculate the ratio of purely exogenous spontaneous emotions relative to video-induced emotions, that is 
$
R^e_0 = \mu^e_0 / \left( \mu^e_0 + \mu^e_1(t) \right)
$
and its complement 
$
R^e_1 = 1 - R^e_0.
$
Figure \ref{fig:ratios} (right) shows the distribution of $\langle R^e_0 \rangle$ versus $\langle R^e_1 \rangle$ over the set of all videos.
The rate of chats induced by videos is typically one-fourth the rate of spontaneous chats.
In summary, this highlights that live chat discussions elicit emotional expressions more effectively through social dynamics compared to passive video content consumption.
In Supplementary Note 5, we further support these findings with additional robustness checks, varying different aspects of our study design, modeling selected subsets of the data sample and emotion categories.
Below, we further show that there is some heterogeneity across video categories, which, however, does not affect the main conclusions of our study.

\subsection*{Emotion Contagion and Interaction Dynamics}
\label{sec:endo-results}

We now interpret the estimated parameters of Equation \eqref{eq:multi_hawkes_general_main} visualized in Figure \ref{fig:result}. 
Here, we focus more on the quantitative aspects of our fit, whereas a qualitative interpretation is given in the following Discussion section. 
Recall that $\alpha^{e,f}$ (Figure \ref{fig:result}, left) represents the rate at which chat-induced emotion $f$ leads to emotion $e$ within the chat (endo). 
Conversely, $\nu^{e,f}$ (Figure \ref{fig:result}, right) signifies the rate at which video-induced emotion $f$ induces emotion $e$ within the chat (exo).
In the preceding section, we observed that most live chat messages originate endogenously, which means that, on average, $\alpha^{e,f}$ significantly exceeds $\nu^{e,f}$. 
Consequently, we shall concentrate our analysis on $\alpha^{e,f}$.

A key quantity is the spectral radius of the branching ratio matrix $\alpha^{e,f}$, denoted $\rho$, which provides a global measure of self-excitation and cross-excitation.  
Mathematically, the spectral radius is defined as the dominant eigenvalue of the matrix $\alpha^{e,f}$, reflecting the overall rate at which events trigger new events.
It captures the extent to which activity is self-sustaining or dependent on external input.
If $\rho > 1$, the process is supercritical, which means that each event generates, on average, more than one new event, leading to an exponential explosion of activity.  
This scenario is generally not realistic, as unbounded growth is unsustainable in most real-world systems, except in rare cases such as nuclear chain reactions.  
In contrast, if $\rho < 1$, the system is subcritical, which means that endogenous activity alone is insufficient to sustain itself indefinitely and new exogenous inputs are required to keep the system active.  
The total expected number of events triggered by a single exogenous event is given by $1/(1 - \rho)$ \cite{laub2021elements}.  
A system at criticality (i.e., with a spectral radius at or very close to $1$) represents a special regime where a single exogenous event can, in principle, generate an indefinitely long cascade of endogenous events.  
For example, it is debated whether brain activity \cite{Maturana2020, Troude2024} and financial transactions operate near criticality \cite{blanc2017quadratic, wehrli2021}.  
Similarly, research on social contagion and information cascades suggests that systems near criticality optimize responsiveness while avoiding runaway diffusion, enabling the rapid spread of ideas and emotions under the right conditions \cite{Accard2019}.

Across our 100 bootstrapped estimates, we estimate the spectral radius to be $0.80 \pm 0.008$, which places the system well below criticality. 
This results in an expected cascade size of $5.0 \pm 0.2$, meaning that each exogenous emotion, on average, generates a cascade of five total events before activity dies out.
Thus, while the system exhibits substantial self-excitation, it remains far from the near-critical regime.  
A similar pattern holds at the level of individual emotions, where the rate of endogeneity for each emotion type $f$ is given by $\sum_e \alpha^{e,f}$.  

Zooming in on the entries of the $\alpha^{e,f}$ matrix reveals that each of the 6 emotions displays distinct temporal dynamics (Figure \ref{fig:result}, left). 
One can easily notice that joy has significantly higher levels of self-excitation rate than other emotions, as shown by the comparison between the different diagonal entries $\alpha^{e,e}$ representing self-excitation.
Joy is up to 3 times more contagious than even high-arousal negative emotions of anger and disgust.
The fact that surprise is the least contagious of all emotions serves as a useful consistency check. 
Similarly, fear shows relatively low levels of excitation.

Generally, self-exciting emotion contagion effects dominate over emotion interaction effects that are quantified by the off-diagonal values $\alpha^{e,f}$ representing the influence of emotion $f$ on emotion $e$.
A notable exception is the moral emotions of anger and disgust, which are mutually reinforcing in almost equal magnitudes. 
It should also be noted that all emotions, including disgust and sadness, trigger joy.
We will come back to these observations in the Discussion section below. 

Recall that the decay times $\gamma^e$ characterize the direct time scale over which past emotions trigger future ones of type $e$. 
Figure \ref{fig:ratios} shows that users have a significantly longer direct memory of negative emotions, in particular high-arousal emotions of anger and disgust.
Coupled with the fact that emotions are triggered primarily through self-excitation, this suggests that negative emotions impose a more lasting direct influence on future interactions.

\subsection*{Variation Across Video Categories}
\label{sec:subsample}

Our main analysis examines general emotional and behavioral patterns among online users by compiling 92,412 YouTube live chat messages across 27 topics.
The susceptibility of emotion dynamics to video influence and peer interactions is potentially subject to the type of video and the audience community.
Here, we conduct sub-sample analyses on representative video types to identify heterogeneity in the evolution of online emotion dynamics.
We focus on 4 video categories with keywords: politics, sports, live, and podcast, with sample sizes 3,043, 11,172, 4,156, and 5,836 live chat messages, respectively.
We report the model parameter estimates and relative ratios of different sources of emotion intensity in Supplementary Note 5.

Across different video categories, we observe several consistent patterns that align with our main findings: joy remains the most contagious emotion and exhibits the highest spontaneous intensity.
Emotion dynamics are largely dominated by endogenous intensities except for the emotions of surprise and fear, which are shown to be more responsive to external stimuli, particularly for sport events and podcast shows. 
We also identify distinct patterns associated with specific video types and audience characteristics. 
The cross-triggering effects of anger and disgust are particularly potent for political videos, which contribute to moral outrage commonly observed in the political context.
Unlike our main result (Figure~\ref{fig:ratios}), the content of political videos has a comparatively larger influence on the emotion dynamics of viewers, consistent with their provocative nature. 
In podcast and live-themed shows, surprise demonstrates a particularly long-lasting direct effect. 
Moreover, surprising content in podcast videos elicits joyful reactions.
These findings validate our methodology by demonstrating its ability to disentangle the distinct channels through which video content and audience characteristics shape emotional dynamics.

In Supplementary Note 5, we show that our results remain robust when using alternative parameterizations of video influence, varying data preprocessing steps, excluding videos that primarily feature interactions between the audience and the video content, or modeling a reduced set of emotions.
In addition, we explicitly compare results between the full dataset and a sub-sample that excludes videos with highly interactive livestreamers.  
Although numerical parameter estimates differ slightly between these two sets, two-sample t-tests reveal no statistically significant differences, indicating that the distributions of estimates are indistinguishable.  
This suggests that feedback from the livestreamer itself plays a negligible role in shaping the overall emotional dynamics observed in our study.

\section*{Discussion} 

\subsection*{Endogenous Influences Dominate Exogenous Influences}

We find that live videos, while central to content delivery, have comparably less direct influence on users' emotional expressions than peer interactions.

Among 92,412 live chat messages on YouTube, emotional dynamics are disproportionately shaped by user interactions rather than passive video consumption, with peer-driven effects outweighing video influence by a factor of four.  
This supports the view that emotions are inherently communicative, emerging primarily through social interaction rather than passive reception \cite{andersen1996principles}.  

The dominance of peer interaction in shaping online emotional expression is consistent with Social Presence Theory \cite{short1976social}, 
which posits that communication channels vary in perceived social presence, influencing interaction patterns.  
This theory, widely applied in research on online learning \cite{lowenthal2010evolution}, suggests that active engagement enhances social presence.  
In live-stream settings, an interactive audience heightens perceived presence in chat discussions, fostering greater emotional expression than passive video viewing. 
Specifically, the unique interactive features of live chats encourage social interactions and community forming which are identified as primary drivers for live-stream engagement \cite{hilvert2018social, hamilton2014streaming}.

Beyond general social influence, emotional mimicry and social appraisal processes further explain the primacy of peer interactions.  
Emotional mimicry depends on contextual interpretation rather than direct replication of expressions \cite{hess2013emotional}.  
On YouTube Live, video content provides the contextual backdrop, but emotions propagate comparably more through user interactions.  
Similarly, emotional expressions function as social signals, helping individuals navigate ambiguity by inferring meaning from others’ emotions \cite{van2015deriving}.  
In this setting, chat-based emotional expressions guide audience engagement with video content.  
This aligns with Affect Theory of Social Exchange \cite{lawler2001affect}, which holds that emotions emerging from social interactions reinforce group cohesion and collective experiences.  

Together, these findings suggest that YouTube Live fosters a socially constructed emotional environment, where peer interactions disproportionately drive emotional expression compared to the video content. 
Consequently, as language models grow more sophisticated, the threat of bots mimicking human users to endogenously manipulate collective emotions becomes an increasing concern \cite{nature2024empathic}.

\subsection*{Positive Emotions Are More Contagious than Negative Emotions}

Our statistical model allows us to disentangle emotional self- and cross-excitation in live chats. 
We find that self-excitation clearly dominates: each emotion is most likely to trigger a recurrence of itself rather than give rise to a different emotion.  
This may help explain why much of the existing research on emotional contagion in online settings has focused on emotions in isolation.  
Nonetheless, non-negligible cross-excitation effects, discussed below, indicate that interactions between emotions also play a role.  

Within self-excitation, we observe a clear hierarchy.
\textit{Joy} exhibits the strongest contagion, followed by moderate levels for \textit{anger}, \textit{disgust}, and \textit{sadness}, and minimal levels for \textit{surprise} and \textit{fear}.  
Specifically, joy’s self-excitation intensity (0.77) is more than three times that of anger and disgust (0.25) and nearly four times that of sadness (0.21).  
This pattern is consistent with prior findings on users’ tendency to express and amplify positive emotions on social platforms \cite{spottswood2016positivity, utz2015function}, and may be rooted in the evolutionary role of positive emotions in fostering social bonding and cooperation \cite{barsade2002ripple}.  
Anger, disgust, and sadness also show meaningful levels of self-excitation, supporting their function as social signals that coordinate group responses to video content \cite{keltner1999social, van2009emotions}.  

By contrast, surprise and fear show little evidence of contagion.  
This is consistent with the fleeting nature of surprise, which often transitions into other emotions following appraisal and sense-making, limiting its transmissibility \cite{noordewier2016temporal}.  
Surprise is also shown to be less contagious in face-to-face interactions \cite{lundqvist1995facial}.  
Fear, likewise, may show limited contagion due to its strong dependence on situational context \cite{ledoux2003emotional}.  

Taken together, this hierarchy of emotional contagion - led by joy - highlights how both the psychological profile and social function of emotions shape their transmissibility in online environments.

\subsection*{Emotions Cross-Excite Each Other}

While emotional contagion has been widely studied, less attention has been given to how different emotions trigger one another in large-scale social interactions.  
Much of the existing literature has focused on how individuals transition between emotional states \cite{kuppens2010feelings}, with some evidence that emotions of similar valence can mutually reinforce one another \cite{pe2012dynamic}.  
In social contexts, one person's emotional expression may elicit not only mimicry but also different emotional responses in others, depending on the interpretation and context \cite{hareli2008emotion}.  
The notion of emotion cycles has been proposed to describe how emotions within groups recursively shape each other over time \cite{hareli2008emotion}, 
and people are known to use multidimensional mental models to predict emotional transitions in others \cite{thornton2017mental}.  

We contribute to this line of research by analyzing the dynamic interplay of six basic emotions in online social interactions.  
Our results reveal two prominent modes of cross-excitation. 

First, anger and disgust strongly reinforce one another.  
Their reciprocal triggering, along with high self-excitation, aligns with the role of these moral emotions in driving moral outrage.  
Previous work has shown that co-occurrence of anger and disgust is predictive of moral outrage \cite{salerno2013}; 
our findings suggest this co-occurrence may be further amplified by mutual excitation.  
This mechanism could help explain the frequent escalation of outrage in online discourse and its potential contribution to polarization and social fragmentation \cite{brady2023overperception, kross2021}.  
We note, however, that part of this effect may be due to the frequent co-labeling of anger and disgust in the same message (see Supplementary Note 1).  

Second, we observe a counterintuitive pattern in which negative emotions such as disgust and sadness trigger joy.
This may reflect \textit{Schadenfreude}, trolling and antisocial behaviors, where users express positive affect in response to others’ negative emotions.  
Online disinhibition and platform affordances may encourage such dynamics, reinforcing gratification from provocation rather than empathy \cite{suler2004online, sanfilippo2018multidimensionality, cheng2017anyone}. 
These findings underscore the need for platform designs that actively encourage prosocial behavior on social media \cite{Dorr2025}.

\subsection*{Positive Emotions are More Contagious, Negative Emotions Last Longer}

Research on emotional contagion in online networks has yielded mixed findings on positivity and negativity biases.  
While self-representation and social validation mechanisms promote positivity \cite{schreurs2021introducing, sheldon2016instagram, spottswood2016positivity},  
negativity often spreads more easily due to cognitive biases that prioritize negative information \cite{rozin2001negativity, knobloch2020confirmation, soroka2019cross}.  
Some studies suggest that positive emotions are more contagious \cite{coviello2014, ferrara2015measuring, gruzd2011happiness},  
whereas others find that negative content tends to reach broader audiences \cite{robertson2023negativity, schone2023negative},  
highlighting the context-dependent nature of emotional virality \cite{paletz2023emotional}.  

Many of these studies, however, do not account for differences in the temporal dynamics and base frequencies of emotional expression.  
By contrast, our Hawkes process framework allows us to disentangle the temporal dependencies governing emotional contagion.  
Our findings indicate that positivity and negativity biases manifest along distinct dimensions.  
Positive emotions are more contagious: a single positive message triggers, on average, 0.77 additional positive messages.  
By comparison, negative emotions such as anger and disgust trigger only 0.26 further negative messages on average.

However, the time scale of these cascades differs markedly.  
Consistent with Negativity Bias Theory \cite{baumeister2001bad},  
negative emotions unfold over a time window nearly twice as long as that of positive emotions,  
suggesting that although less contagious in volume, negative affect persists longer in online conversations. 
These findings resonate with previous observations that positive emotions rise quickly and fade faster,
while negative emotions accumulate gradually before dissipating \cite{naskar2020, fan2019, pellert2020}.

\coloredtext{
\subsection*{Limitations}
}

A key limitation of our analysis is that we approximate the emotional content of the video solely through its subtitles.  
While this text-based approach captures the semantic cues of emotional expression, it omits prosodic and visual signals such as tone of voice, facial expressions, and body language.  
Previous studies have shown that discrete emotion measures derived from text correlate with self-reported emotions, but may diverge from facial and vocal indicators \cite{munin2025language}.  
Future work integrating multimodal data, including audio and visual cues, could improve emotion inference and offer a more complete picture of how emotional dynamics unfold across communication channels.  

Another limitation is that we do not distinguish between human users and automated accounts (bots) in YouTube live chats.  
This distinction would enable a more refined application of our framework, allowing the analysis of emotion dynamics separately for human and automated agents.  
Given the increasing prevalence of bots and the rapid development of large language models capable of simulating emotional expression,  
future studies should examine how emotional exchanges evolve in mixed human-AI environments \cite{stella2018bots, ishowo2019behavioural, oudah2024}.  

Another limitation of the present study is that emotions are modeled as discrete categories of equal magnitude. 
Although based on the widely used six basic emotions model \citep{ekman1992}, it does not account for variations in emotional intensity or the continuous structure of affective experience. 
An immediate extension would be to adopt a \textit{marked Hawkes process}, in which each emotional event is associated with a continuous-valued mark reflecting its strength or salience \citep{reinhart2018review}. 
This could be achieved by learning to assign each emotional expression a continuous intensity score, or, presumably even more insightfully, by adopting a continuous representation of emotions grounded in the \textit{circumplex model of affect}.
In this model, emotions are embedded in a two- or three-dimensional space defined by axes such as valence (positive vs. negative affect), arousal (high vs. low activation), and dominance (control vs. submission) \citep{Russell1980,Mehrabian1996}. 
A marked Hawkes process defined over such a space would enable the modeling of how emotional expressions not only evolve in frequency but also drift through affective space over time. 
This would allow us to capture more fine-grained patterns of emotional contagion, such as whether high arousal is more contagious than low arousal, or whether emotional dynamics differ along the dominance axis.

In parallel, future work could relax the assumption of additivity inherent in classical Hawkes models.
One particularly promising extension is the \textit{Quadratic Hawkes process} which incorporates both linear and quadratic feedback terms \citep{blanc2017quadratic,kanazawa2023asymptotic}.
This allows the intensity of future emotional expressions to depend not only on the sum of past events, but also on pairwise combinations of past emotions. 
Such a framework naturally accommodates non-linear interactions, e.g., whether the co-occurrence of surprise and sadness increases the likelihood of subsequent anger, and even allows for inhibition, as negative-valued quadratic terms can suppress future intensities. 
By combining the continuous circumplex representation of emotions with a marked Quadratic Hawkes process, future research could investigate how different dimensions of affect interact to shape emotional dynamics at the group level in real time.

Another promising direction for future work is to reframe the definition of what is considered endogenous and exogenous while applying exactly the same modeling framework.  
In our current study, we treat the entire audience collectively as the endogenous system, with video content providing the primary exogenous input.  
Alternatively, one could define each individual viewer as the endogenous component and treat all other viewers and the video as exogenous influences.  
This reframing would not require any changes to the methodology itself: the same Hawkes process framework and estimation procedures can be directly applied, simply interpreting the resulting parameters at the individual rather than group level.  
Such an approach would allow for a richer understanding of how much an individual's emotional expressions are shaped by their own prior states versus external social and media influences.

Together, these extensions offer a powerful and flexible modeling framework to capture the rich, non-linear, and multi-dimensional nature of emotion contagion in online and offline interactions.

\section*{Conclusions} 

We analyzed 92,412 YouTube Live chat messages from 397 videos to study emotional contagion in social networks and livestream platforms. 
Our findings show that user interactions are four times more emotionally influential than passive video content. 
Joy is three times more contagious than negative emotions, but negative emotions persist nearly twice as long, potentially creating lasting negative moods. 
We also observed cross-excitation effects, where negative posts can trigger positive emotions in trolls, revealing antisocial dynamics that harm community cohesion. These insights have practical implications: promoting joyful content and enabling active communication (e.g., live chats or bots) can foster emotional engagement and positivity. 
However, the lingering impact of negative emotions highlights the need for timely moderation. 
Platforms must balance emotional amplification with community well-being. 
Designing for positive contagion while managing toxic interactions is key to sustaining healthy online environments. 
Future research could explore cultural influences and the long-term impacts of emotional contagion strategies.


\section*{Data Availability Statement} 
Publicly available data are obtained from YouTube.
Data needed to replicate this study are available at \url{https://github.com/ivylyis/Quantification-of-the-Self-Excited-Emotion-Dynamics-in-Online-Interactions}.

\section*{Code Availability Statement} 

Code needed to replicate this study, in Python 3.7.6, is available at \url{https://github.com/ivylyis/Quantification-of-the-Self-Excited-Emotion-Dynamics-in-Online-Interactions}.


{
\bibliographystyle{unsrt}
\bibliography{bibliography.bib}
}

\coloredtext{
\section*{Acknowledgements}
The authors received no specific funding for this work.
}
\section*{Author Contributions}
S.C.L., D.S., and Y.L. designed the research; 
Y.L. aggregated the data and wrote the code;
S.C.L. D.S., and Y.L. analyzed the results; 
and S.C.L., D.S., and Y.L. wrote and reviewed the manuscript.

\section*{Competing Interest}
The authors declare no competing interests.

\onecolumn
\appendix
\begin{center}
    {\fontsize{18}{22}\selectfont \bfseries Supplementary Information for} \\
    \vspace{2em}
    {\fontsize{14}{22}\selectfont \bfseries Social feedback amplifies emotional language in online video live chats}
\end{center}

\begin{appendices} 
\setcounter{section}{0}                      
\renewcommand{\thesection}{\arabic{section}} 

\titleformat{\section}
  {\normalfont\Large\bfseries}               
  {Supplementary Note \thesection:}          
  {1em}                                      
  {}                                         

\makeatletter
\renewcommand{\theHsection}{suppnote.\arabic{section}}
\makeatother

\setcounter{figure}{0}                
\renewcommand{\thefigure}{\arabic{figure}}  

\setcounter{table}{0}                 
\renewcommand{\thetable}{\arabic{table}}

\renewcommand{\figurename}{Supplementary Figure}
\renewcommand{\tablename}{Supplementary Table}


\section{Summary Statistics of YouTube Live Chat Data} 
\label{sec:apx_livechat}

We show some additional summary statistics of our dataset. 
Once we filtered the data by removing inactive videos, we compiled 92,412 live chat messages from 397 videos, amassing a total of 780 hours of material and encompassing 27 distinct topics.
The distribution of topics is shown in  Supplementary Figure \ref{fig:keyword}. 
The distribution of emotional events is shown in Supplementary Figure \ref{fig:livechat_descriptive}.
We also show the distribution of the median time between each event arrival in Supplementary Figure \ref{fig:livechat_descriptive_interval}. 
To exemplify our emotion labeling approach, we showcase representative live chat messages and predicted emotions in Supplementary Table \ref{tab:livechat_examples}.

Since we label emotions non-exclusively, we show the co-occurrence patterns across 6 basic emotions in Supplementary Figure \ref{fig:emotion_cooccur}. 
In general, negative emotions have a higher tendency to co-occur with one another.
In particular, anger and disgust have a relatively high co-occurrence.

To examine whether expressing different emotions requires systematically different levels of effort, we analyze the distribution of message lengths across emotion categories. 
As shown in Supplementary Figure~\ref{fig:text_len}, the variation in text length within each category is substantially larger than the differences between categories. 
This suggests that observed asymmetries in emotional dynamics between positive and negative emotions are unlikely to be driven by differences in text length or typing effort.

\begin{figure*}[!htb]
	\centering
	\includegraphics[width=\textwidth]{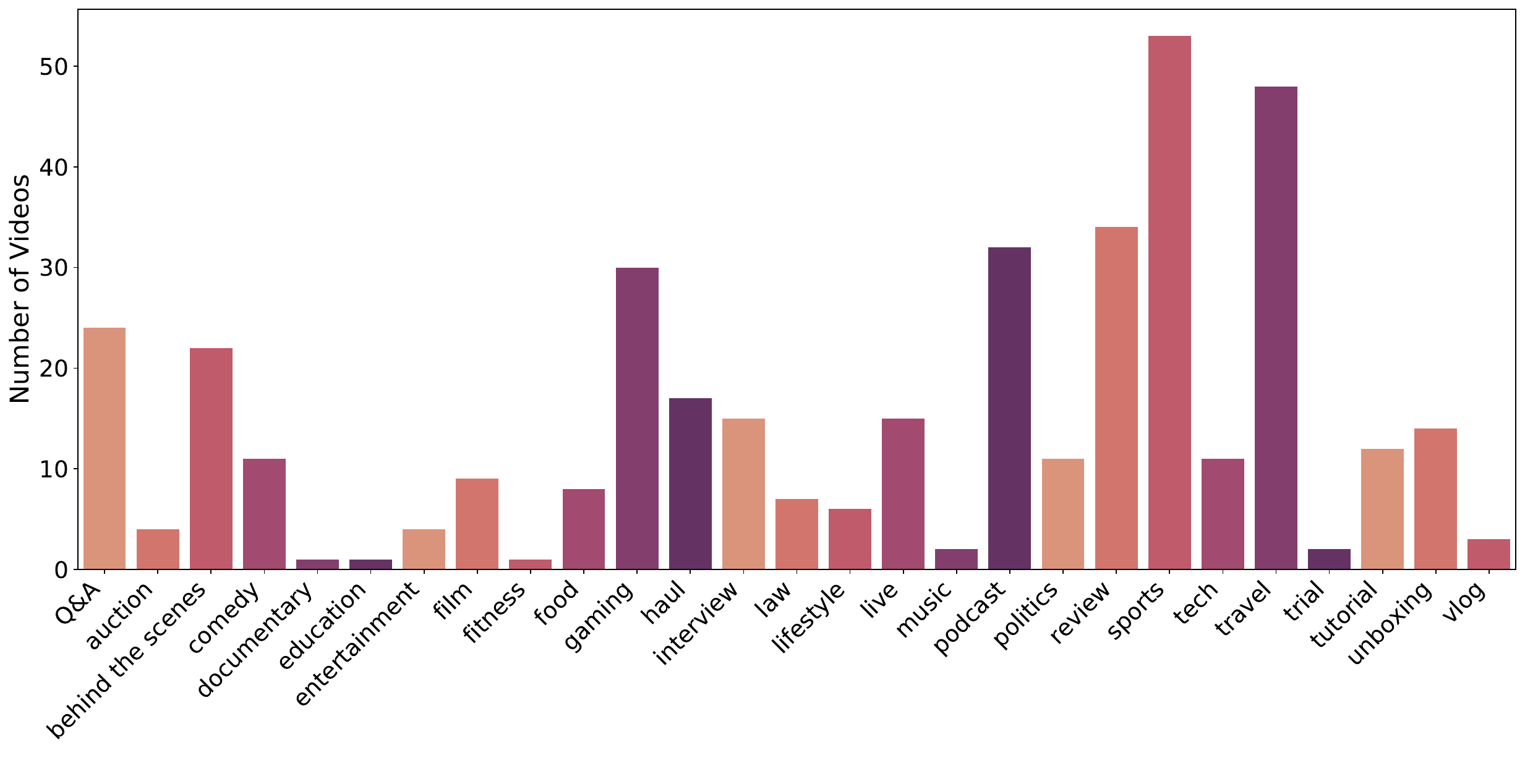}
	\caption{
			Number of videos per keyword in our final data sample across 27 keywords. 
		}
	\label{fig:keyword}
\end{figure*}

\begin{figure*}[!htb]
	\centering
	\includegraphics[width=\textwidth]{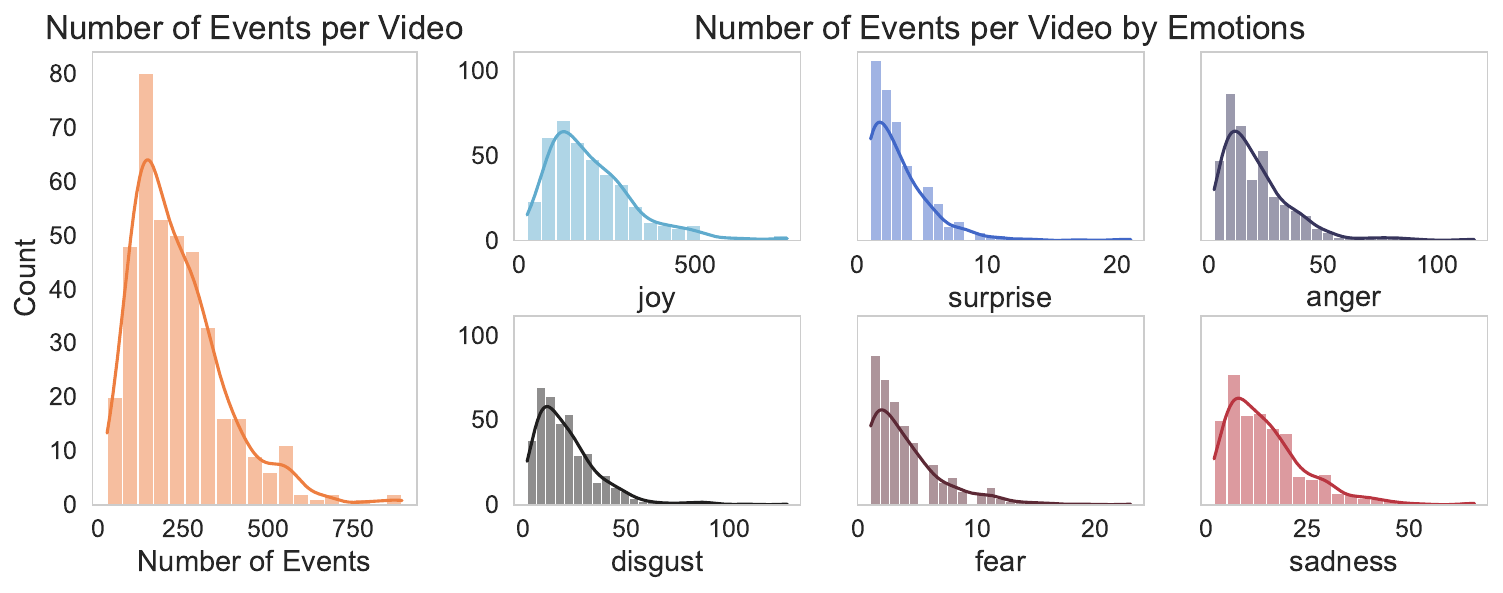}
	\caption{
		(left) Distribution of the number of events per video across 397 videos. 
            (right) Same as left but broken down per emotion. 
		}
	\label{fig:livechat_descriptive}
\end{figure*}

\begin{figure*}[!htb]
	\centering
	\includegraphics[width=\textwidth]{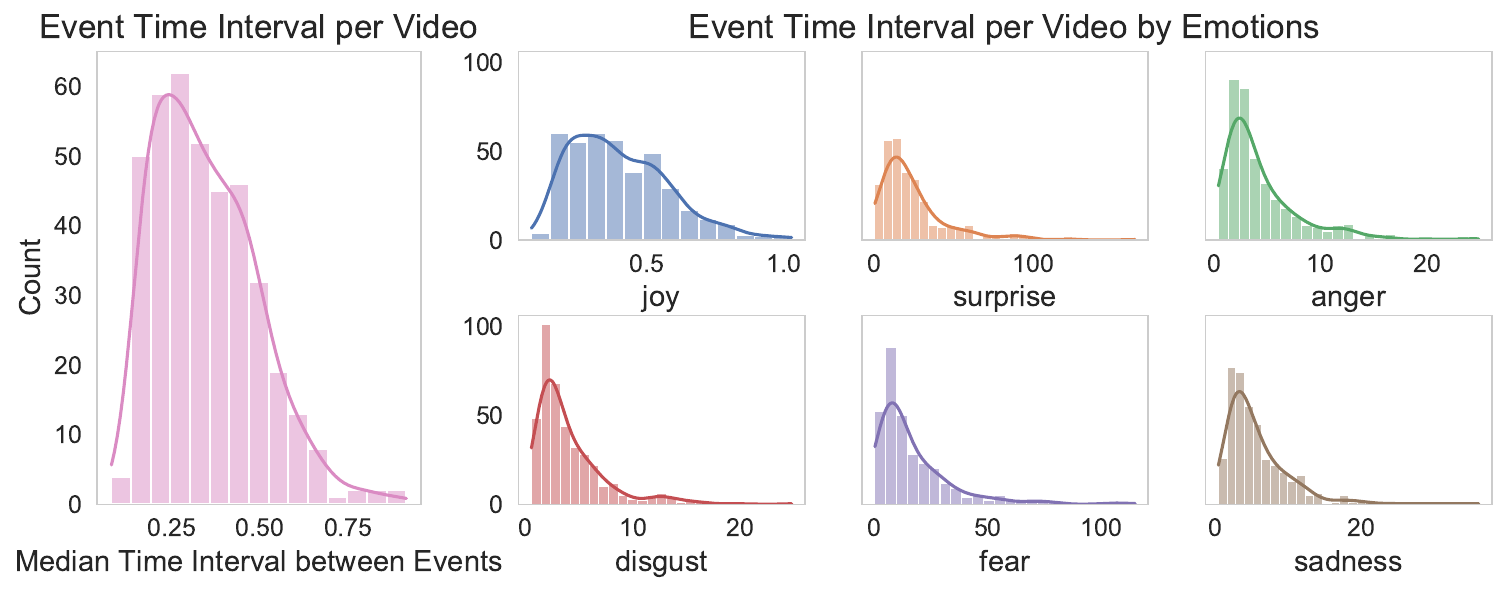}
	\caption{
			(left) Distribution of the median time interval between each event in units of minutes for each video across 397 videos.
            (right) Same as left but broken down per emotion. 
		}
	\label{fig:livechat_descriptive_interval}
\end{figure*}

\begin{figure*}[!htb]
	\centering
	\includegraphics[width=0.7\textwidth]{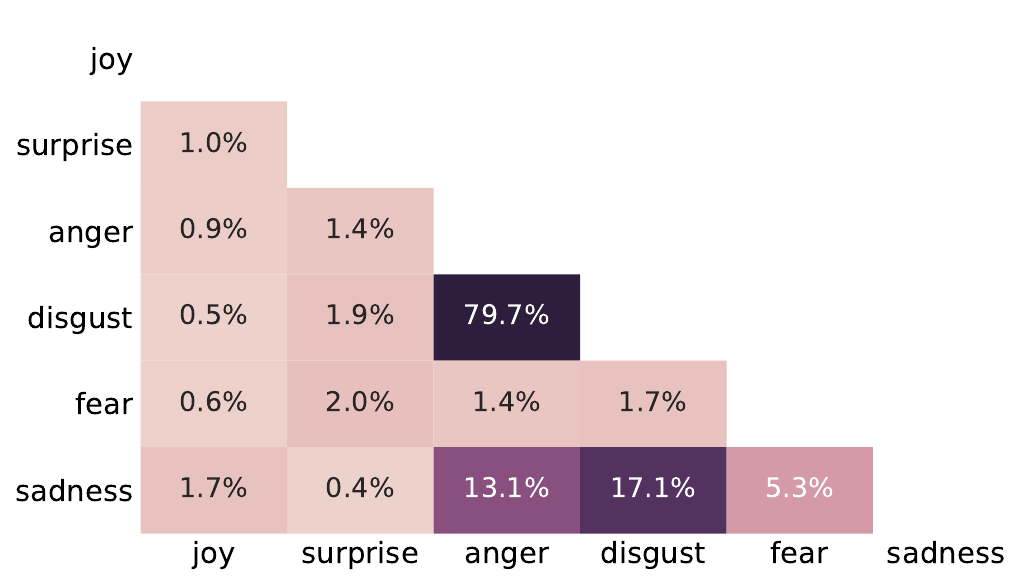}
	\caption{
		Co-occurrence of emotions in our data. 
        We normalize the number of co-occurrences for each emotion pair by the sum of occurrences of the two emotions in the live chat.
   }
	\label{fig:emotion_cooccur}
\end{figure*}

\begin{figure}[!htb]
	\centering
	\includegraphics[width=0.8\textwidth]{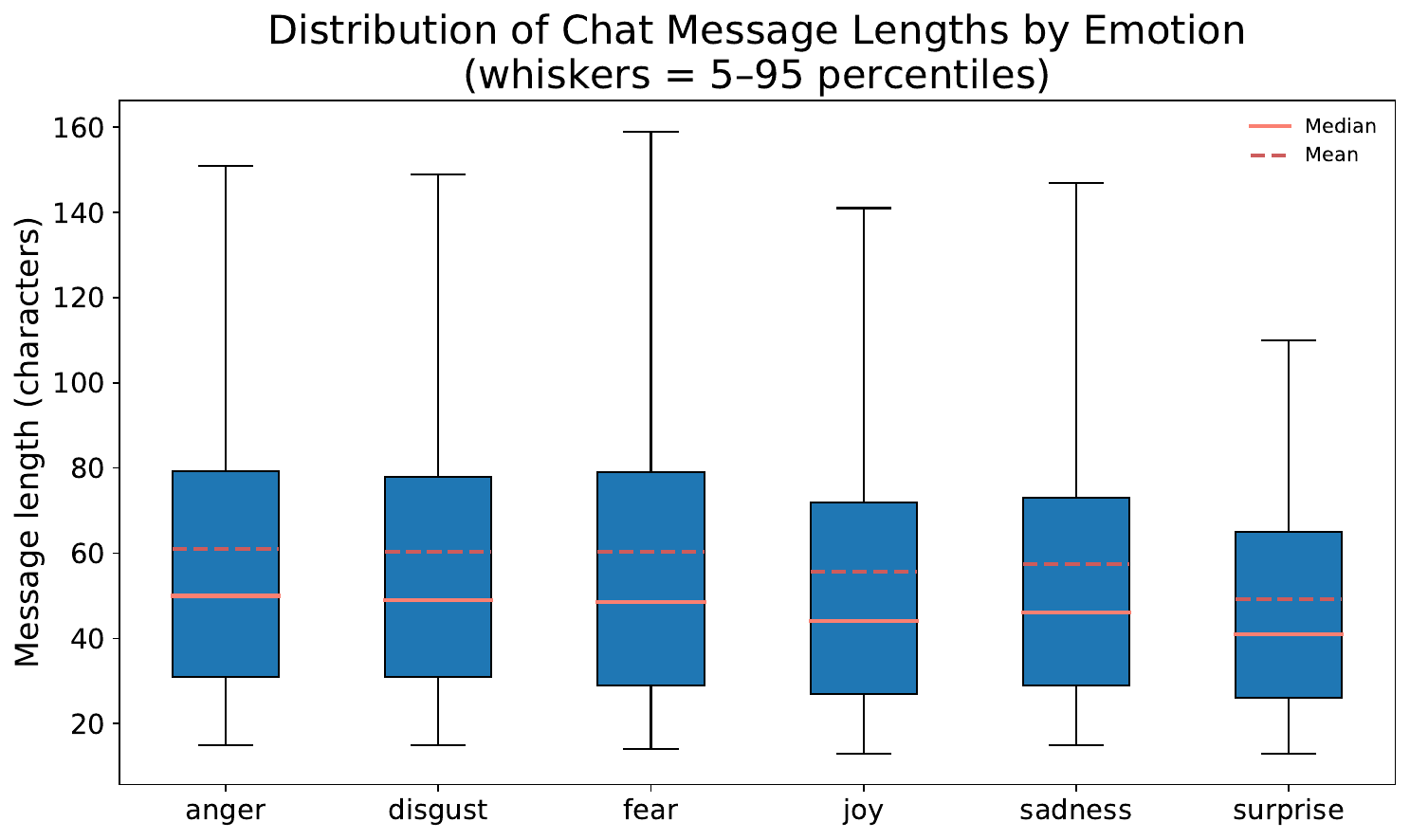}
	\caption{
Distribution of YouTube Live Chat message lengths across emotion categories. 
The whiskers indicate the 5th to 95th percentiles, with both the median and mean marked within each box. 
The variation in text length within each emotion category is substantially larger than the differences in average text length across categories, 
suggesting that message length is not systematically biased by emotion type.
    }
	\label{fig:text_len}
\end{figure}

\begin{table*}[ht]
\centering
\caption{Sample live chat messages and detected emotions}
\label{tab:livechat_examples}
\begin{tabular}{p{10cm}l}
\toprule
\textbf{Live-chat comment} & \textbf{Emotion(s)}\\
\midrule
Thanks so much for doing this. I feel like I got to go on an adventure during my lunch break! & joy \\
\midrule
That is unbelievable & surprise \\
\midrule
I hate laugh tracks!! Some of the shows on Netflix use it and it drives me nuts! & anger, disgust \\
\midrule
I can't stand haters on high horses reporting people that are not even doing anything morally wrong & anger, disgust \\
\midrule
His voice sounds scary & fear \\
\midrule
So sorry, don't cry anymore & sadness \\
\midrule
What time is it & no emotions \\
\bottomrule
\end{tabular}
\end{table*}

\clearpage
\section{Baseline Intensity Calibration in Hawkes Process} 
\label{sec:baseline_intensity}

Correctly calibrating the baseline intensity is crucial to the reliability of Hawkes process estimations.
Insights from the field of finance illustrate how non-constant background intensity can bias the branching ratio estimation \cite{filimonovSor2015,wheatley2019, wehrli2021}.
There has been a debate about whether financial markets are operating at criticality.
Calibration of the Hawkes model in financial time series had shown that the branching ratio is less than 1, thus rejecting criticality \cite{filimonov2012}.
Subsequently, an alternative estimation reported a branching ratio close to 1, suggesting that markets operate at criticality \cite{hardiman2013}.
Both studies obtained their results from the calibration of a temporal Hawkes model assuming constant baseline intensity.
The opposing findings have to do with the length of the financial time series and the non-stationarity.
Specifically, the latter study used time series of long durations over multiple months \cite{hardiman2013}.
In contrast, the former study rejecting criticality used much shorter time series from 10 to 30 minutes due to the concern of nonstationarity
of longer time series \cite{filimonov2012}.
As shown by subsequent studies, financial time series are in fact temporally nonstationary, highly volatile at market openings and closings, and stable around lunch hours \cite{wheatley2019, wehrli2021}.
As a result, calibrating the Hawkes model with time series over multiple days with the erroneous assumption of constant baseline intensity will artificially inflate the branching ratio towards 1 to account for the non-stationarity omitted in the background.
Using an advanced Hawkes model calibration that employs the expectation maximization (EM) method to parameterize non-constant baseline intensity over financial time series spanning several months, the hypothesis of market criticality is rejected \cite{wheatley2019, wehrli2021}.

In the field of earthquake prediction, advanced Epidemic-Type Aftershock Sequences (ETAS) (also known as Hawkes model in the domain of finance and mathematics) calibration has been proposed which features a superior parametric representation of the spatially varying background seismicity rate \cite{nandan2021, nandan2022}.
A recent study further improves the methodology to avoid biases in the estimation of the branching ratio \cite{li2024revisiting}. 
Studies have shown that the branching ratio of earthquake activities on the global level, for New Zealand, California, as well as Yunnan and Sichuan provinces of China are below 1, revealing subcritical seismicity \cite{nandan2021, nandan2022, li2024revisiting}.
However, it has been shown that constraining to a constant baseline intensity results in the false conclusion that the branching ratio approaches 1, suggesting seismicity criticality \cite{nandan2021, nandan2022, li2024revisiting}.
Assuming a constant baseline intensity in the ETAS estimation neglects the spatial heterogeneity of the background seismicity.
Consequently, the branching ratio is biased towards 1 where the perceived criticality is a statistical artifact arising from the incorrectly specified ETAS model attempting to account for the activities in the background seismicity.

In conclusion, it is essential to allow for nonstationarity and nonuniformity of the baseline intensity in order to avoid several biases
in the parameter estimation of the Hawkes process, and in particular in the determination of its branching ratio.

\clearpage
\section{Log-Likelihood Derivation}
\label{sec:likelihood_derivation}

We define a general multivariate Hawkes self-excited conditional point process \cite{saichev2011generating,saichev2013hierarchy} as follows.
We observe $|\mathcal{E}|$ point processes.
For each process $e \in \mathcal{E}$, we observe $N^{e}$ events, $i = 1, \ldots N^{e}$, where event $i$ of process $e$ takes place at time point $t_i^{e}$, $i = 1, \ldots N^{e}$.
We define $t_0^{e} \equiv 0$, $t_{N^{e}+1}^{e} \equiv T$ for $e \in \mathcal{E}$, such that the observation period throughout the process is $\{ t_i^{e} \mid i^{e} = 0, \ldots, N^{e}+1 \}$.
We denote the intensity at time $t$ for a given process as
\begin{equation}
	\lambda^{e}(t)
	=  \mu^{e}(t) + \sum_{f \in \mathcal{E}} \sum_{t_j^{f} < t} \phi^{e, f}(t-t_j^{f}),
	\label{eq:multi_hawkes_general}
\end{equation} 
where $t_j^{f}$ enumerates all past events of process $f$ that took place before $t$.
The function $\phi^{e,f}(t)$ is defined to only take positive values.
The term $\mu^{e}(t)$ represents the exogenous component of the Hawkes process leading to spontaneous occurrence of events, 
whereas $\sum_{f \in \mathcal{E}} \sum_{t_j^{f} < t} \phi^{e, f}(t-t_j^{f})$ represents the endogenous impact from past events in process $f$ on process $e$.
In the current form, we can estimate all parameters for fixed $e$. 

We derive the log-likelihood function for a fairly general class of the Hawkes process as follows. 
The expected number of events for emotion $e$ within time interval $\Delta t$ is approximately $\lambda^e(t) \Delta t$ for small $\Delta t$'s.
In the limit of an infinitesimally small time interval $ \Delta t \to \mathrm{d}t$, 
the term $\lambda^e(t) \mathrm{d}t$ then represents the probability of observing exactly one event, 
since the probability of observing more than one event is an infinitesimal of higher order  (technically proportional to $dt^2$). 
Similarly, the probability of not observing any event within $\mathrm{d}t$ reads $1-\lambda^e(t) \mathrm{d}t$, 
which, in the limit of $dt$ infinitesimal, becomes equivalent to $\exp \left(-\lambda^e(t) \mathrm{d}t \right)$. 
The probability of not observing any event over a finite time $\Delta t$ is thus 
$
\exp \left( - \nt{t}{t+\Delta t}{\tau}{\lambda^e(\tau)} \right)
$
by the law of independent probabilities. 
Taken together, the likelihood function capturing event intensity at $t_i$ over $[0, T]$ reads 
\begin{equation}
	\prod_{i=1}^{N^{e}}\lambda^e(t_i) e^{-\nt{t_i}{t_{i+1}}{s}{\lambda^e(s)}}.
	\label{eq:hawkes_lh}
\end{equation} 
Taking the product over the exponential terms, Equation \eqref{eq:hawkes_lh} simplifies to $\prod_{i=1}^{N^{e}} \lambda^e(t_i) e^{ -\nt{0}{T}{s}{\lambda^e(s)}}$.
The Hawkes process likelihood function is thus 
\begin{equation}
	L^{e}( \theta^{e, f}) 
        = \prod_{i=1}^{N^{e}}\lambda^{e}(t_i) e^{ -\nt{0}{T}{s}{\lambda^{e}(s)}}.
	\label{eq:multi_lh}
\end{equation} 
The log-likelihood function of a general multivariate Hawkes process is thus
\begin{subequations}
\label{eq:multi_hawkes_llh}
\begin{align}
	\log(L^{e}(\theta^{e, f}))
        &= \sum_{i=1}^{N^{e}} \log(\lambda^{e}(t_i)) -\nt{0}{T}{s}{\lambda^{e}(s)} \\
        &= \sum_{i=1}^{N^{e}}\log(\lambda^{e}(t_i)) - \nt{0}{T}{s}{\mu^{e}(s)} -  \sum_{f \in \mathcal{E}} \sum_{t_j^{f}}\nt{t_j^{f}}{T}{s}{\phi^{e, f}(s)}.
	\label{eq:multi_llh_general}
\end{align}
\end{subequations}
It is common to assume $\phi^{e, f}(t) = \alpha^{e, f} \frac{1}{\gamma^e} e^{-\frac{1}{\gamma^e} t}$.
For simplicity and as a way to address the fact that the parameter $\gamma^e$ is ``sloppy'', which means that the log-likelihood function 
has a very large radius of curvature close to its maximum along the $\gamma^e$ dimension, we assume $\gamma^e$ to be constant across all processes.
Expanding Equation \eqref{eq:multi_llh_general}, the log-likelihood function is written as
\begin{subequations}
\begin{align}
        \log L^{e}(\alpha^{e, f}, \gamma^e)
        &\overset{\eqref{eq:multi_llh_general}}{=}\sum_{i=1}^{N^{e}}\log(\lambda^{e}(t_i)) - \nt{0}{T}{s}{\mu^{e}(s)} - \sum_{f \in \mathcal{E}} \sum_{t_j^{f}}\nt{t_j^{f}}{T}{s}{\alpha^{e, f} \frac{1}{\gamma^e} e^{-\frac{1}{\gamma^e} (s - t_j^{f})}} \\ 
        &= \sum_{i=1}^{N^{e}}\log(\lambda^{e}(t_i)) - \nt{0}{T}{s}{\mu^{e}(s)} + \sum_{f \in \mathcal{E}}\sum_{t_j^{f}}{\alpha^{e, f} e^{-\frac{1}{\gamma^e} (s - t_j^{f})}}\Big|_{t_j^{f}}^T \\ 
        &= \sum_{i=1}^{N^{e}}\log \left(\mu^{e}(t_i) + \sum_{f \in \mathcal{E}} \sum_{t_j^{f} < t} \alpha^{e, f} \frac{1}{\gamma^e} e^{-\frac{1}{\gamma^e} (t_i - t_j^{f})} \right) - \nt{0}{T}{s}{\mu^{e}(s)}  +  \notag \\
        &\quad \sum_{f \in \mathcal{E}} \sum_{t_j^{f}} \alpha^{e, f} \left(e^{-\frac{1}{\gamma^e} (T - t_j^{f})} - 1 \right).
        \label{eq:multi_hawkes_llh_expanded}
\end{align}   
\end{subequations}
The following sections involve variations of the baseline intensity function, whereas the endogenous cross-excitation component is unchanged.
We thus define $\sum_{f \in \mathcal{E}} \sum_{t_j^{f}}\alpha^{e, f} \left(e^{-\frac{1}{\gamma^e} (T - t_j^{f})} - 1 \right)$ from Equation \eqref{eq:multi_hawkes_llh_expanded} as $endo$ for the following derivation.

In our model, we allow for a coefficient term $\nu^{e, f}$ which scales the effect of emotion $f$ displayed in the video on emotion $e$ in the chat. 
Concretely, we model the cross-excitation effects of the baseline function for process $f$ on process $e$, namely
$\mu^{e}(t) =  \mu^e_0 + \mu^e_1(t) \equiv \mu^e_0 +  \sum_{f \in \mathcal{E}}~ \nu^{e, f} ~S^f(t)$.
We thus denote a multivariate Hawkes process for a given process as
\begin{equation}
	\lambda^{e}(t)
	= \mu^{e}_0 + \sum_{f \in \mathcal{E}}\nu^{e,f}S^{f}(t) + \sum_{f \in \mathcal{E}} \sum_{t_j^{f} < t} \phi^{e, f}(t-t_j^{f})
	\label{eq:multi_hawkes_general_extend}
\end{equation} 
where $S^{f}(t)$ is known.
The log-likelihood function is thus
\begin{subequations}
\label{eq:multi_hawkes_llh_corss_mu}
\begin{align}
	\log(L^{e}(\mu^{e}_0, \nu^{e, f}, \theta^{e, f}))
        &= \sum_{i=1}^{N^{e}} \log(\lambda^{e}(t_i)) -\nt{0}{T}{s}{\lambda^{e}(s)} \\
        &= \sum_{i=1}^{N^{e}}\log(\lambda^{e}(t_i)) - \nt{0}{T}{s}{\mu^{e}_0} - \sum_{f \in \mathcal{E}}\nu^{e, f}\nt{0}{T}{s}{S^{f}(s)} -  \sum_{f \in \mathcal{E}} \sum_{t_j^{f}}\nt{t_j^{f}}{T}{s}{\phi^{e, f}(s)}.
	\label{eq:multi_llh_general_cross_mu}
\end{align}
\end{subequations}

We assume $\phi^{e, f}(t) = \alpha^{e, f} \frac{1}{\gamma^e} e^{-\frac{1}{\gamma^e} t}$.
Expanding Equation \eqref{eq:multi_llh_general_cross_mu}, the log-likelihood function is written as
\begin{subequations}
\begin{align}
        \log L^{e}(\mu^{e}_0, \nu^{e, f}, \alpha^{e, f}, \gamma^e)
        &\overset{\eqref{eq:multi_llh_general_cross_mu}}{=}\sum_{i=1}^{N^{e}}\log(\lambda^{e}(t_i)) - \nt{0}{T}{s}{\mu^{e}_0} - \sum_{f \in \mathcal{E}}\nu^{e, f}\nt{0}{T}{s}{S^{f}(s)} + endo \\ 
        &= \sum_{i=1}^{N^{e}}\log(\lambda^{e}(t_i)) -  \nt{0}{T}{s}{\mu^{e}_0} - \sum_{f \in \mathcal{E}}\nu^{e, f}\nt{0}{T}{s}{S^{f}(s)} + endo \\ 
        &= \sum_{i=1}^{N^{e}}\log \left(\mu^{e}_0 + \sum_{f \in \mathcal{E}}\nu^{e,f}S^{f}(t) + \sum_{f \in \mathcal{E}} \sum_{t_j^{f} < t} \alpha^{e, f} \frac{1}{\gamma^e} e^{-\frac{1}{\gamma^e} (t_i - t_j^{f})} \right) - \notag \\
        & \quad \nt{0}{T}{s}{\mu^{e}_0} - \sum_{f \in \mathcal{E}}\nu^{e, f}\nt{0}{T}{s}{S^{f}(s)} + endo 
        \label{eq:hawkes_llh_expanded_cross_mu}
\end{align}   
\end{subequations}
We compute $\nt{0}{T}{s}{S^{f}(s)}$ numerically as $M^f_1$. Without loss of generality, we simplify the log-likelihood function to 
\begin{align} \label{apx_eq:likelihood}
	\log L^{e}(\mu^{e}_0, \nu^{e, f}, \alpha^{e, f}, \gamma^e) 
        &= \sum_{i=1}^{N^{e}}\log \left( \mu^{e}_0 + \sum_{f \in \mathcal{E}}\nu^{e, f}S^{f}(t_i) + \sum_{f \in \mathcal{E}} \sum_{t_j^{f} < t} \alpha^{e, f} \frac{1}{\gamma^e} e^{-\frac{1}{\gamma^e} (t_i - t_j^{f})} \right) - \notag \\
        & \quad \mu^{e}_0 T - \sum_{f \in \mathcal{E}}\nu^{e, f}M^f_1 + \sum_{f \in \mathcal{E}} \sum_{t_j^{f}}\alpha^{e, f} \left(e^{-\frac{1}{\gamma^e} (T - t_j^{f})} - 1 \right).
\end{align}

\clearpage
\section{Testing Log-Likelihood Fits with Synthetic Data}
\label{sec:synthetic_data}

We estimate parameter values that maximize the log-likelihood function \eqref{apx_eq:likelihood} via Quasi-Newton optimization. 
Without loss of generality, we normalize the log-likelihood function by the number of events $N^{e}$, in order to compare the log-likelihood values across different sample sizes.
We use the Python \textit{ticks} library 
(\url{https://x-datainitiative.github.io/tick/index.html})
to simulate Hawkes processes with parameters that we refer to as the ground-truth.
Subsequently, we compare the estimated parameters with their ground-truth values to evaluate performance.
Since the parameters for each emotion are estimated separately, without loss of generality, we only show the synthetic test results for one emotion (process).
The other processes have comparable performances.
To assess the robustness of the estimation, we bootstrap the data and repeat the estimation 100 times.

\begin{figure*}[!htb]
	\centering
	\includegraphics[width=\textwidth]{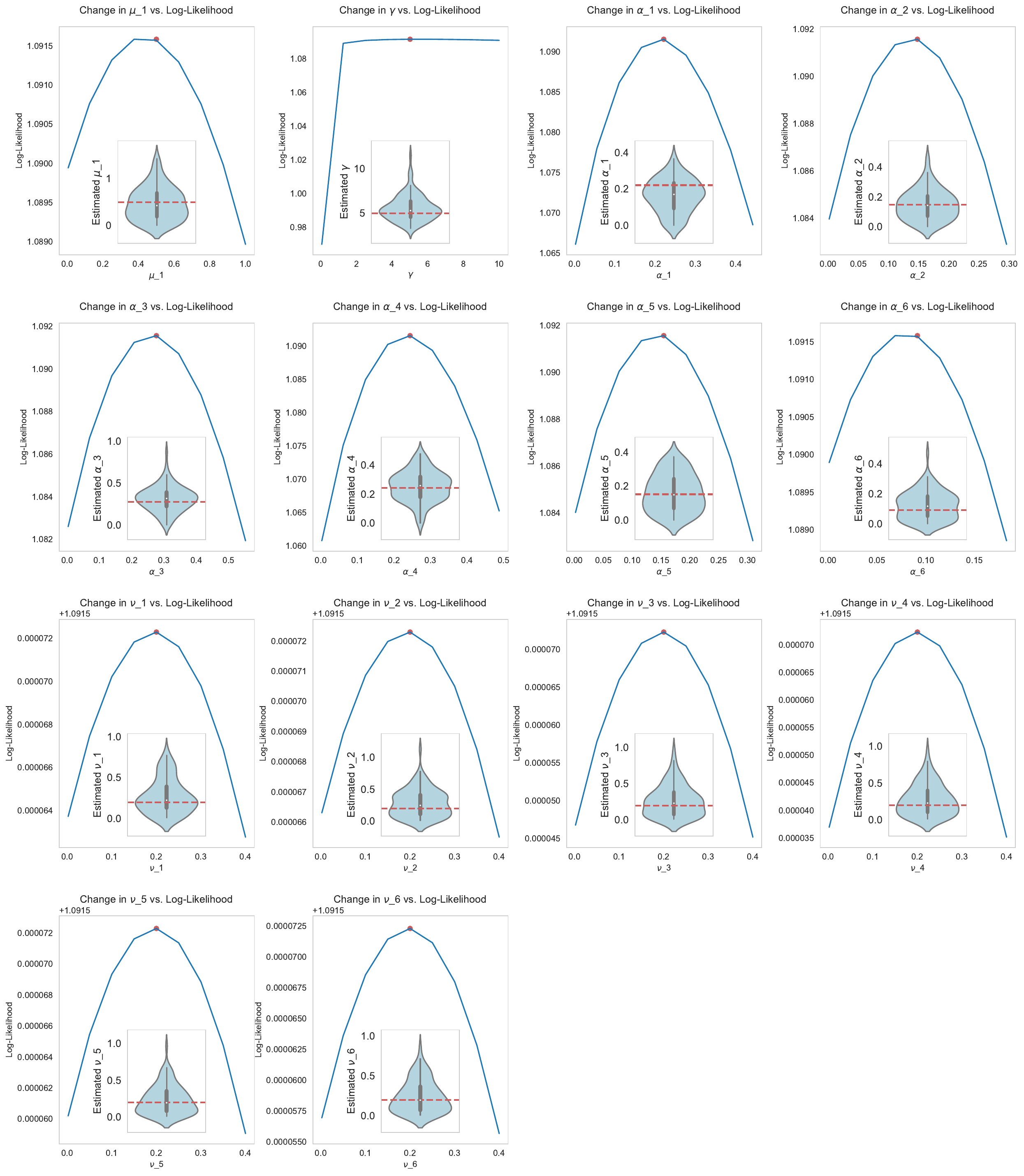}
	\caption{
	        We plot the shape of the log-likelihood function in Equation \eqref{apx_eq:likelihood} varying the target parameter while holding the other parameters fixed at their ground-truth values. 
	        The ground-truth values of the target parameter are indicated by the red dots.
                The log-likelihood values are normalized by the number of events $N^{e}$ to be comparable across varying numbers of data points.
                The inset plots show the distribution of the fitted parameters from 100 bootstrapped samples. 
                The red dashed lines indicate the ground-truth values for the parameter of interest.
                We show results for the first process in a 6-dimensional multivariate Hawkes process.
                The other processes have comparable performances.
			}
	\label{fig:llh_baseline_cross_excitation}
\end{figure*}

The inset plots of Supplementary Figure \ref{fig:llh_baseline_cross_excitation} shows the distributions of estimated $\mu^{1}_0$, $\gamma^1$, $\alpha^{1, f}$, and $\nu^{1, f}$ values for 100 bootstrap samples with $N^{e}$ around 10,000.
The dashed red lines mark the ground-truth values for $\mu^{1}_0$, $\gamma^1$, $\alpha^{1, f}$, and $\nu^{1, f}$ respectively.
The estimated parameters are reasonably robust and are centered on the ground-truth values.
In addition, we plot the log-likelihood functions with respect to changes in the parameter value.
We observe that the log-likelihood functions are well-behaved and maximized by the ground-truth parameter values, with the exception of $\gamma$.
The value of the logarithmic likelihood function is insensitive to changes in $\gamma^1$ over a certain range, displaying sloppiness \cite{machta2013parameter}. 

\clearpage
\section{Robustness Checks} 
\label{sec:robust_add}
In the following sections, we present a series of robustness checks that support the validity of our main findings.  
We demonstrate that our results remain qualitatively consistent under alternative parameterizations of video influence, variations in data preprocessing, the exclusion of specific video types, and the modeling of a reduced set of emotions.
In addition, we complement our main analysis with a study of non-emotional livechat dynamics.
These additional analyses reinforce the reliability of our conclusions and highlight the robustness of the findings across modeling choices.

\subsection{Comparison with Null Distribution} 
\label{sec:null}

To assess whether the parameters estimated by our Hawkes model capture meaningful excitation dynamics rather than statistical noise, we conduct a null distribution analysis in the following. 
The null hypothesis posits that there is no self- or cross-excitation of emotions, meaning no emotional contagion or interaction occurs in the YouTube Live setting.
Under this null, all entries of the $\boldsymbol{\alpha}$ and $\boldsymbol{\nu}$ matrices are set to zero.
To simulate this scenario, we generated synthetic data from a six-dimensional Hawkes process in which the ground-truth $\boldsymbol{\alpha}$ and $\boldsymbol{\nu}$ matrices are all zero.
In this setting, each emotion is independently driven by a constant baseline exogenous rate, $\mu_0^e = 0.5$ for all $e$.
We generated 100 synthetic datasets under this null specification, following the procedure described in Supplementary Note \ref{sec:synthetic_data} above.
We then re-estimated the model parameters for each null dataset using the same log-likelihood maximization procedure as described in the Methods section.
This yielded a distribution of parameter estimates that represents the null distribution, which effectively describes the expected behavior of the model in the absence of emotional excitation and video-driven influence.

We compare the distributions of the $\alpha$ and $\nu$ parameters estimated from our empirical data with their corresponding null distributions across all six basic emotions to assess whether the observed effects are statistically significant.  
Specifically, we apply the two-sided Kolmogorov--Smirnov (KS) test to each parameter to examine whether the distributions differ with statistical significance.
The quasi-zero $p$-values allow us to reject the null hypothesis for all estimated parameters, demonstrating statistically meaningful excitation effects.   
We obtain qualitatively similar results under the scenario where the ground-truth $\boldsymbol{\alpha}$ are all zero and $\boldsymbol{\nu}$ are non-zero.
Notably, for certain $\nu$ parameters, the empirical estimates consistently reached the lower boundary of $10^{-6}$.
Consequently, we indicate the p-values of the KS test as NaN. 
This is because of the technical challenges in interpreting $p$-values associated with point distributions, and also due to the evident visible difference.

Overall, this null distribution analysis provides strong evidence that our estimated parameters are significantly different from the null and reflect meaningful self- and cross-excitation dynamics in emotional expression.

\subsection{Robustness Check on Video Influence} 
\label{sec:video_robustness}

\begin{table*}[!htb]
\centering
\large
\setlength{\tabcolsep}{5pt}
\renewcommand{\arraystretch}{1.12}
\caption{
Mean estimates of endogenous excitation parameters $\alpha_{e,f}$ and decay rates $\gamma^{e}$ (left), and exogenous influence parameters $\nu_{e,f}$ and spontaneous baselines $\mu_0^{e}$ (right) for different model specfications. 
The \textit{Baseline} model corresponds to our main analysis, with results also reported in Figure 2 in the main text.  
For all other models, we re-estimate the parameters using subsets of videos filtered by the specified keywords.  
}
\label{tab:param_keyword}
\begin{minipage}[t]{0.49\textwidth}
\centering
\subcaption*{\textbf{(a) Endogenous $\alpha_{e,f}$ and $\gamma^{e}$}}
\begin{adjustbox}{max width=\linewidth}
\begin{tabular}{lccccc}
\toprule
\textbf{Parameter} & \textbf{Baseline} & \textbf{Politics} & \textbf{Sports} & \textbf{Live} & \textbf{Podcast}\\
\midrule
\multicolumn{5}{l}{\textbf{\emph{Joy}}}\\
$\alpha_{\mathrm{joy,joy}}$       & 0.76 & 0.62 & 0.63 & 0.74 & 0.73\\
$\alpha_{\mathrm{joy,surprise}}$  & 0.03 & 0.07 & 0.00 & 0.92 & 0.00\\
$\alpha_{\mathrm{joy,anger}}$     & 0.00 & 0.01 & 0.14 & 0.00 & 0.17\\
$\alpha_{\mathrm{joy,disgust}}$   & 0.20 & 0.35 & 0.11 & 0.17 & 0.01\\
$\alpha_{\mathrm{joy,fear}}$      & 0.08 & 0.05 & 0.07 & 0.35 & 0.00\\
$\alpha_{\mathrm{joy,sadness}}$   & 0.21 & 0.19 & 0.24 & 0.20 & 0.22\\
$\gamma^{\text{joy}}$             & 2.36 & 3.34 & 2.87 & 2.45 & 3.88\\
\midrule
\multicolumn{5}{l}{\textbf{\emph{Surprise}}}\\
$\alpha_{\mathrm{surprise,joy}}$       & 0.01 & 0.00 & 0.00 & 0.01 & 0.00\\
$\alpha_{\mathrm{surprise,surprise}}$  & 0.05 & 0.05 & 0.07 & 0.22 & 0.00\\
$\alpha_{\mathrm{surprise,anger}}$     & 0.01 & 0.00 & 0.00 & 0.00 & 0.00\\
$\alpha_{\mathrm{surprise,disgust}}$   & 0.01 & 0.02 & 0.01 & 0.01 & 0.01\\
$\alpha_{\mathrm{surprise,fear}}$      & 0.01 & 0.00 & 0.01 & 0.02 & 0.00\\
$\alpha_{\mathrm{surprise,sadness}}$   & 0.01 & 0.03 & 0.02 & 0.01 & 0.08\\
$\gamma^{\text{surprise}}$             & 3.44 & 7.53 & 2.61 & 10.86 & 14.54\\
\midrule
\multicolumn{5}{l}{\textbf{\emph{Anger}}}\\
$\alpha_{\mathrm{anger,joy}}$       & 0.02 & 0.04 & 0.02 & 0.02 & 0.04\\
$\alpha_{\mathrm{anger,surprise}}$  & 0.07 & 0.24 & 0.05 & 0.04 & 0.05\\
$\alpha_{\mathrm{anger,anger}}$     & 0.26 & 0.22 & 0.29 & 0.21 & 0.27\\
$\alpha_{\mathrm{anger,disgust}}$   & 0.20 & 0.34 & 0.10 & 0.27 & 0.01\\
$\alpha_{\mathrm{anger,fear}}$      & 0.10 & 0.09 & 0.07 & 0.21 & 0.20\\
$\alpha_{\mathrm{anger,sadness}}$   & 0.04 & 0.00 & 0.02 & 0.06 & 0.10\\
$\gamma^{\text{anger}}$             & 4.81 & 3.81 & 3.42 & 7.66 & 3.91\\
\midrule
\multicolumn{5}{l}{\textbf{\emph{Disgust}}}\\
$\alpha_{\mathrm{disgust,joy}}$       & 0.02 & 0.05 & 0.02 & 0.04 & 0.05\\
$\alpha_{\mathrm{disgust,surprise}}$  & 0.05 & 0.24 & 0.05 & 0.00 & 0.05\\
$\alpha_{\mathrm{disgust,anger}}$     & 0.20 & 0.26 & 0.29 & 0.22 & 0.26\\
$\alpha_{\mathrm{disgust,disgust}}$   & 0.26 & 0.30 & 0.10 & 0.27 & 0.01\\
$\alpha_{\mathrm{disgust,fear}}$      & 0.09 & 0.10 & 0.07 & 0.19 & 0.21\\
$\alpha_{\mathrm{disgust,sadness}}$   & 0.06 & 0.00 & 0.02 & 0.03 & 0.08\\
$\gamma^{\text{disgust}}$             & 4.64 & 3.63 & 3.42 & 6.41 & 3.85\\
\midrule
\multicolumn{5}{l}{\textbf{\emph{Fear}}}\\
$\alpha_{\mathrm{fear,joy}}$       & 0.01 & 0.01 & 0.00 & 0.01 & 0.01\\
$\alpha_{\mathrm{fear,surprise}}$  & 0.02 & 0.06 & 0.02 & 0.03 & 0.00\\
$\alpha_{\mathrm{fear,anger}}$     & 0.02 & 0.00 & 0.02 & 0.00 & 0.00\\
$\alpha_{\mathrm{fear,disgust}}$   & 0.00 & 0.00 & 0.00 & 0.08 & 0.00\\
$\alpha_{\mathrm{fear,fear}}$      & 0.12 & 0.15 & 0.08 & 0.30 & 0.12\\
$\alpha_{\mathrm{fear,sadness}}$   & 0.04 & 0.03 & 0.05 & 0.01 & 0.03\\
$\gamma^{\text{fear}}$             & 3.33 & 8.77 & 6.17 & 4.36 & 2.71\\
\midrule
\multicolumn{5}{l}{\textbf{\emph{Sadness}}}\\
$\alpha_{\mathrm{sadness,joy}}$       & 0.03 & 0.04 & 0.02 & 0.06 & 0.03\\
$\alpha_{\mathrm{sadness,surprise}}$  & 0.06 & 0.11 & 0.06 & 0.03 & 0.00\\
$\alpha_{\mathrm{sadness,anger}}$     & 0.01 & 0.00 & 0.01 & 0.12 & 0.00\\
$\alpha_{\mathrm{sadness,disgust}}$   & 0.07 & 0.06 & 0.06 & 0.04 & 0.05\\
$\alpha_{\mathrm{sadness,fear}}$      & 0.11 & 0.00 & 0.20 & 0.08 & 0.12\\
$\alpha_{\mathrm{sadness,sadness}}$   & 0.21 & 0.06 & 0.16 & 0.12 & 0.14\\
$\gamma^{\text{sadness}}$             & 3.61 & 2.38 & 2.77 & 4.15 & 2.74\\
\bottomrule
\end{tabular}
\end{adjustbox}
\end{minipage}
\hfill
\begin{minipage}[t]{0.49\textwidth}
\centering
\subcaption*{\textbf{(b) Exogenous $\nu_{e,f}$ and $\mu_{0}^{e}$}}
\begin{adjustbox}{max width=\linewidth}
\begin{tabular}{lccccc}
\toprule
\textbf{Parameter} & \textbf{Baseline} & \textbf{Politics} & \textbf{Sports} & \textbf{Live} & \textbf{Podcast}\\
\midrule
\multicolumn{5}{l}{\textbf{\emph{Joy}}}\\
$\nu_{\mathrm{joy,joy}}$          & 0.03 & 0.11 & 0.01 & 0.00 & 0.02\\
$\nu_{\mathrm{joy,surprise}}$     & 0.05 & 0.52 & 0.01 & 0.15 & 0.05\\
$\nu_{\mathrm{joy,anger}}$        & 0.00 & 0.12 & 0.00 & 0.01 & 0.00\\
$\nu_{\mathrm{joy,disgust}}$      & 0.00 & 0.00 & 0.00 & 0.00 & 0.00\\
$\nu_{\mathrm{joy,fear}}$         & 0.00 & 0.04 & 0.02 & 0.00 & 0.02\\
$\nu_{\mathrm{joy,sadness}}$      & 0.00 & 0.11 & 0.00 & 0.01 & 0.00\\
$\mu_{0}^{\text{joy}}$            & 0.29 & 0.10 & 0.50 & 0.32 & 0.35\\
\midrule
\multicolumn{5}{l}{\textbf{\emph{Surprise}}}\\
$\nu_{\mathrm{surprise,joy}}$     & 0.00 & 0.00 & 0.00 & 0.00 & 0.00\\
$\nu_{\mathrm{surprise,surprise}}$& 0.00 & 0.09 & 0.06 & 0.00 & 0.00\\
$\nu_{\mathrm{surprise,anger}}$   & 0.00 & 0.01 & 0.00 & 0.02 & 0.00\\
$\nu_{\mathrm{surprise,disgust}}$ & 0.00 & 0.00 & 0.00 & 0.00 & 0.00\\
$\nu_{\mathrm{surprise,fear}}$    & 0.00 & 0.01 & 0.01 & 0.01 & 0.00\\
$\nu_{\mathrm{surprise,sadness}}$ & 0.00 & 0.00 & 0.00 & 0.00 & 0.00\\
$\mu_{0}^{\text{surprise}}$       & 0.01 & 0.00 & 0.01 & 0.01 & 0.01\\
\midrule
\multicolumn{5}{l}{\textbf{\emph{Anger}}}\\
$\nu_{\mathrm{anger,joy}}$        & 0.00 & 0.01 & 0.00 & 0.00 & 0.00\\
$\nu_{\mathrm{anger,surprise}}$   & 0.02 & 0.02 & 0.01 & 0.00 & 0.00\\
$\nu_{\mathrm{anger,anger}}$      & 0.03 & 0.08 & 0.02 & 0.00 & 0.01\\
$\nu_{\mathrm{anger,disgust}}$    & 0.01 & 0.00 & 0.00 & 0.04 & 0.01\\
$\nu_{\mathrm{anger,fear}}$       & 0.01 & 0.05 & 0.03 & 0.00 & 0.00\\
$\nu_{\mathrm{anger,sadness}}$    & 0.00 & 0.03 & 0.00 & 0.03 & 0.00\\
$\mu_{0}^{\text{anger}}$          & 0.04 & 0.02 & 0.05 & 0.02 & 0.05\\
\midrule
\multicolumn{5}{l}{\textbf{\emph{Disgust}}}\\
$\nu_{\mathrm{disgust,joy}}$      & 0.00 & 0.00 & 0.00 & 0.00 & 0.00\\
$\nu_{\mathrm{disgust,surprise}}$ & 0.03 & 0.00 & 0.02 & 0.01 & 0.02\\
$\nu_{\mathrm{disgust,anger}}$    & 0.02 & 0.09 & 0.02 & 0.00 & 0.01\\
$\nu_{\mathrm{disgust,disgust}}$  & 0.01 & 0.00 & 0.00 & 0.04 & 0.00\\
$\nu_{\mathrm{disgust,fear}}$     & 0.02 & 0.03 & 0.03 & 0.00 & 0.00\\
$\nu_{\mathrm{disgust,sadness}}$  & 0.00 & 0.05 & 0.00 & 0.02 & 0.00\\
$\mu_{0}^{\text{disgust}}$        & 0.04 & 0.03 & 0.06 & 0.02 & 0.04\\
\midrule
\multicolumn{5}{l}{\textbf{\emph{Fear}}}\\
$\nu_{\mathrm{fear,joy}}$         & 0.00 & 0.00 & 0.00 & 0.00 & 0.00\\
$\nu_{\mathrm{fear,surprise}}$    & 0.01 & 0.00 & 0.00 & 0.00 & 0.04\\
$\nu_{\mathrm{fear,anger}}$       & 0.00 & 0.00 & 0.00 & 0.00 & 0.01\\
$\nu_{\mathrm{fear,disgust}}$     & 0.00 & 0.01 & 0.00 & 0.01 & 0.00\\
$\nu_{\mathrm{fear,fear}}$        & 0.03 & 0.01 & 0.03 & 0.00 & 0.05\\
$\nu_{\mathrm{fear,sadness}}$     & 0.01 & 0.01 & 0.01 & 0.00 & 0.00\\
$\mu_{0}^{\text{fear}}$           & 0.01 & 0.00 & 0.01 & 0.00 & 0.01\\
\midrule
\multicolumn{5}{l}{\textbf{\emph{Sadness}}}\\
$\nu_{\mathrm{sadness,joy}}$      & 0.00 & 0.00 & 0.00 & 0.00 & 0.00\\
$\nu_{\mathrm{sadness,surprise}}$ & 0.00 & 0.23 & 0.01 & 0.00 & 0.00\\
$\nu_{\mathrm{sadness,anger}}$    & 0.00 & 0.01 & 0.00 & 0.00 & 0.00\\
$\nu_{\mathrm{sadness,disgust}}$  & 0.00 & 0.00 & 0.00 & 0.01 & 0.00\\
$\nu_{\mathrm{sadness,fear}}$     & 0.01 & 0.01 & 0.02 & 0.01 & 0.00\\
$\nu_{\mathrm{sadness,sadness}}$  & 0.02 & 0.04 & 0.00 & 0.01 & 0.01\\
$\mu_{0}^{\text{sadness}}$        & 0.03 & 0.00 & 0.04 & 0.00 & 0.03\\
\bottomrule
\end{tabular}
\end{adjustbox}
\end{minipage}

\end{table*}

\begin{table*}[!htb]
\centering
\caption{
Statistics of emotion dynamics for different model specifications. 
The spectral radius $\rho$ of the branching ratio matrix $\alpha^{e,f}$ provides a global measure of self-excitation and cross-excitation.
For each video, we calculate the average ratio of endogenous (exogenous) intensity to the total intensity across time, and obtain a fraction of endogenous versus exogenous intensity.
For each emotion, we calculate and report the median endogenous vs. exogenous intensity ratio from the distribution of these ratios across videos. 
We obtain the median statistics of spontaneous vs. video-influenced intensity ratios for each emotion in the same manner.
The \textit{Baseline} model corresponds to our main analysis, with results also reported in Figure 3 in the main text.  
For all other models, we report the results using subsets of videos filtered by the specified keywords.  
We notice that emotions are predominantly triggered endogenously. 
}
\label{tab:ratios_keyword}
\large
\setlength{\tabcolsep}{6pt}
\renewcommand{\arraystretch}{1.3}
\begin{tabular}{lccccc}
\toprule
\textbf{Parameter} & \textbf{Baseline} & \textbf{Politics} & \textbf{Sports} & \textbf{Live} & \textbf{Podcast}\\
\midrule
\textbf{\textit{Spectral radius}} & 0.80 & 0.75 & 0.66 & 0.81 & 0.76\\
\midrule
\multicolumn{5}{l}{\textbf{\textit{Endogenous vs.\ Exogenous Intensity}}} \\
\quad joy       & 3.05 & 2.56 & 1.78 & 3.63  & 2.28 \\
\quad surprise  & 1.07 & 2.83 & 0.55 & 1.42  & 0.60 \\
\quad anger     & 1.80 & 3.73 & 1.32 & 3.49  & 1.79 \\
\quad disgust   & 1.93 & 3.33 & 1.13 & 5.01  & 2.29 \\
\quad fear      & 1.77 & 5.19 & 0.77 & 6.79  & 0.79 \\
\quad sadness   & 2.71 & 4.13 & 1.55 & 19.54 & 1.91 \\
\midrule
\multicolumn{5}{l}{\textbf{\textit{Spontaneous Expression vs.\ Video-Induced}}} \\
\quad joy       & 4.08 & 0.89 & 6.50 & 47.25 & 6.33 \\
\quad surprise  & 2.41 & 0.83 & 1.29 & 1.28  & 23.22 \\
\quad anger     & 4.17 & 0.94 & 5.85 & 1.68  & 5.83 \\
\quad disgust   & 4.51 & 1.10 & 6.24 & 1.57  & 12.67 \\
\quad fear      & 2.26 & 0.58 & 0.82 & 1.14  & 1.18 \\
\quad sadness   & 6.08 & 0.44 & 19.10 & 2.23 & 8.76 \\
\bottomrule
\end{tabular}
\end{table*}

\begin{table*}[!htb]
\centering
\large
\setlength{\tabcolsep}{6pt}
\renewcommand{\arraystretch}{1.15}
\caption{
Mean estimates of endogenous excitation parameters $\alpha_{e,f}$ and decay rates $\gamma^{e}$ (left), and exogenous influence parameters $\nu_{e,f}$ and spontaneous baselines $\mu_0^{e}$ (right) for different model specfications. 
The \textit{Baseline} model corresponds to our main analysis, with results also reported in Figure 2 in the main text.  
Columns Factor 0.5 and Factor 2 show the estimated parameters when multiplying the maximum and median in the log-normal function by a factor of 0.5 and 2, respectively. 
Column Power Law shows the estimated parameters when parametrizing the video influence with a power law shape.
The estimated parameters across different models are highly consistent with our main result.
}
\label{tab:param_vid}
\begin{minipage}[t]{0.49\textwidth}
\centering
\subcaption*{\textbf{(a) Endogenous $\alpha_{e,f}$ and $\gamma^{e}$}}
\begin{adjustbox}{max width=\linewidth}
\begin{tabular}{lcccc}
\toprule
\textbf{Parameter} & \textbf{Baseline} & \textbf{Factor 0.5} & \textbf{Factor 2} & \textbf{Power Law} \\
\midrule
\multicolumn{5}{l}{\textbf{\emph{Joy}}}\\
$\alpha_{\mathrm{joy,\,joy}}$       & 0.76 & 0.76 & 0.76 & 0.77\\
$\alpha_{\mathrm{joy,\,surprise}}$  & 0.03 & 0.04 & 0.01 & 0.02\\
$\alpha_{\mathrm{joy,\,anger}}$     & 0.00 & 0.01 & 0.01 & 0.00\\
$\alpha_{\mathrm{joy,\,disgust}}$   & 0.20 & 0.20 & 0.20 & 0.20\\
$\alpha_{\mathrm{joy,\,fear}}$      & 0.08 & 0.08 & 0.08 & 0.06\\
$\alpha_{\mathrm{joy,\,sadness}}$   & 0.21 & 0.21 & 0.21 & 0.21\\
$\gamma^{\mathrm{joy}}$             & 2.36 & 2.36 & 2.32 & 2.42\\
\midrule
\multicolumn{5}{l}{\textbf{\emph{Surprise}}}\\
$\alpha_{\mathrm{surprise,\,joy}}$      & 0.01 & 0.01 & 0.01 & 0.01\\
$\alpha_{\mathrm{surprise,\,surprise}}$ & 0.05 & 0.06 & 0.05 & 0.05\\
$\alpha_{\mathrm{surprise,\,anger}}$    & 0.01 & 0.00 & 0.01 & 0.01\\
$\alpha_{\mathrm{surprise,\,disgust}}$  & 0.01 & 0.01 & 0.01 & 0.00\\
$\alpha_{\mathrm{surprise,\,fear}}$     & 0.01 & 0.01 & 0.01 & 0.01\\
$\alpha_{\mathrm{surprise,\,sadness}}$  & 0.01 & 0.01 & 0.01 & 0.01\\
$\gamma^{\mathrm{surprise}}$            & 3.44 & 3.15 & 3.16 & 2.99\\
\midrule
\multicolumn{5}{l}{\textbf{\emph{Anger}}}\\
$\alpha_{\mathrm{anger,\,joy}}$       & 0.02 & 0.02 & 0.02 & 0.02\\
$\alpha_{\mathrm{anger,\,surprise}}$  & 0.07 & 0.09 & 0.07 & 0.05\\
$\alpha_{\mathrm{anger,\,anger}}$     & 0.26 & 0.27 & 0.24 & 0.25\\
$\alpha_{\mathrm{anger,\,disgust}}$   & 0.20 & 0.20 & 0.22 & 0.20\\
$\alpha_{\mathrm{anger,\,fear}}$      & 0.10 & 0.09 & 0.10 & 0.08\\
$\alpha_{\mathrm{anger,\,sadness}}$   & 0.04 & 0.04 & 0.03 & 0.04\\
$\gamma^{\mathrm{anger}}$             & 4.81 & 4.88 & 4.83 & 4.47\\
\midrule
\multicolumn{5}{l}{\textbf{\emph{Disgust}}}\\
$\alpha_{\mathrm{disgust,\,joy}}$       & 0.02 & 0.02 & 0.02 & 0.02\\
$\alpha_{\mathrm{disgust,\,surprise}}$  & 0.05 & 0.06 & 0.06 & 0.04\\
$\alpha_{\mathrm{disgust,\,anger}}$     & 0.20 & 0.21 & 0.18 & 0.19\\
$\alpha_{\mathrm{disgust,\,disgust}}$   & 0.26 & 0.25 & 0.28 & 0.25\\
$\alpha_{\mathrm{disgust,\,fear}}$      & 0.09 & 0.09 & 0.09 & 0.08\\
$\alpha_{\mathrm{disgust,\,sadness}}$   & 0.06 & 0.06 & 0.06 & 0.06\\
$\gamma^{\mathrm{disgust}}$             & 4.64 & 4.70 & 4.63 & 4.39\\
\midrule
\multicolumn{5}{l}{\textbf{\emph{Fear}}}\\
$\alpha_{\mathrm{fear,\,joy}}$        & 0.01 & 0.01 & 0.00 & 0.01\\
$\alpha_{\mathrm{fear,\,surprise}}$   & 0.02 & 0.02 & 0.02 & 0.02\\
$\alpha_{\mathrm{fear,\,anger}}$      & 0.02 & 0.02 & 0.02 & 0.02\\
$\alpha_{\mathrm{fear,\,disgust}}$    & 0.00 & 0.00 & 0.00 & 0.00\\
$\alpha_{\mathrm{fear,\,fear}}$       & 0.12 & 0.12 & 0.11 & 0.12\\
$\alpha_{\mathrm{fear,\,sadness}}$    & 0.04 & 0.04 & 0.04 & 0.04\\
$\gamma^{\mathrm{fear}}$              & 3.33 & 3.15 & 3.20 & 3.08\\
\midrule
\multicolumn{5}{l}{\textbf{\emph{Sadness}}}\\
$\alpha_{\mathrm{sadness,\,joy}}$       & 0.03 & 0.03 & 0.03 & 0.03\\
$\alpha_{\mathrm{sadness,\,surprise}}$  & 0.06 & 0.05 & 0.06 & 0.05\\
$\alpha_{\mathrm{sadness,\,anger}}$     & 0.01 & 0.01 & 0.01 & 0.01\\
$\alpha_{\mathrm{sadness,\,disgust}}$   & 0.07 & 0.08 & 0.08 & 0.08\\
$\alpha_{\mathrm{sadness,\,fear}}$      & 0.11 & 0.10 & 0.11 & 0.11\\
$\alpha_{\mathrm{sadness,\,sadness}}$   & 0.21 & 0.21 & 0.20 & 0.21\\
$\gamma^{\mathrm{sadness}}$             & 3.61 & 3.54 & 3.59 & 3.56\\
\bottomrule
\end{tabular}
\end{adjustbox}
\end{minipage}
\hfill
\begin{minipage}[t]{0.49\textwidth}
\centering
\subcaption*{\textbf{(b) Exogenous $\nu_{e,f}$ and $\mu_0^{e}$}}
\begin{adjustbox}{max width=\linewidth}
\begin{tabular}{lcccc}
\toprule
\textbf{Parameter} & \textbf{Baseline} & \textbf{Factor 0.5} & \textbf{Factor 2} & \textbf{Power Law} \\
\midrule
\multicolumn{5}{l}{\textbf{\emph{Joy}}}\\
$\nu_{\mathrm{joy,\,joy}}$       & 0.03 & 0.06 & 0.02 & 0.01\\
$\nu_{\mathrm{joy,\,surprise}}$  & 0.05 & 0.18 & 0.03 & 0.09\\
$\nu_{\mathrm{joy,\,anger}}$     & 0.00 & 0.01 & 0.00 & 0.00\\
$\nu_{\mathrm{joy,\,disgust}}$   & 0.00 & 0.00 & 0.00 & 0.00\\
$\nu_{\mathrm{joy,\,fear}}$      & 0.00 & 0.04 & 0.00 & 0.00\\
$\nu_{\mathrm{joy,\,sadness}}$   & 0.00 & 0.03 & 0.00 & 0.00\\
$\mu_0^{\mathrm{joy}}$           & 0.29 & 0.29 & 0.29 & 0.32\\
\midrule
\multicolumn{5}{l}{\textbf{\emph{Surprise}}}\\
$\nu_{\mathrm{surprise,\,joy}}$       & 0.00 & 0.00 & 0.00 & 0.00\\
$\nu_{\mathrm{surprise,\,surprise}}$  & 0.00 & 0.00 & 0.01 & 0.02\\
$\nu_{\mathrm{surprise,\,anger}}$     & 0.00 & 0.00 & 0.00 & 0.00\\
$\nu_{\mathrm{surprise,\,disgust}}$   & 0.00 & 0.00 & 0.00 & 0.00\\
$\nu_{\mathrm{surprise,\,fear}}$      & 0.00 & 0.00 & 0.00 & 0.00\\
$\nu_{\mathrm{surprise,\,sadness}}$   & 0.00 & 0.01 & 0.00 & 0.01\\
$\mu_0^{\mathrm{surprise}}$           & 0.01 & 0.01 & 0.01 & 0.01\\
\midrule
\multicolumn{5}{l}{\textbf{\emph{Anger}}}\\
$\nu_{\mathrm{anger,\,joy}}$       & 0.00 & 0.00 & 0.00 & 0.00\\
$\nu_{\mathrm{anger,\,surprise}}$  & 0.02 & 0.06 & 0.01 & 0.03\\
$\nu_{\mathrm{anger,\,anger}}$     & 0.03 & 0.05 & 0.01 & 0.05\\
$\nu_{\mathrm{anger,\,disgust}}$   & 0.01 & 0.00 & 0.02 & 0.00\\
$\nu_{\mathrm{anger,\,fear}}$      & 0.01 & 0.04 & 0.00 & 0.00\\
$\nu_{\mathrm{anger,\,sadness}}$   & 0.00 & 0.01 & 0.00 & 0.00\\
$\mu_0^{\mathrm{anger}}$           & 0.04 & 0.04 & 0.04 & 0.04\\
\midrule
\multicolumn{5}{l}{\textbf{\emph{Disgust}}}\\
$\nu_{\mathrm{disgust,\,joy}}$       & 0.00 & 0.00 & 0.00 & 0.00\\
$\nu_{\mathrm{disgust,\,surprise}}$  & 0.03 & 0.06 & 0.02 & 0.04\\
$\nu_{\mathrm{disgust,\,anger}}$     & 0.02 & 0.04 & 0.01 & 0.03\\
$\nu_{\mathrm{disgust,\,disgust}}$   & 0.01 & 0.00 & 0.02 & 0.01\\
$\nu_{\mathrm{disgust,\,fear}}$      & 0.02 & 0.04 & 0.00 & 0.00\\
$\nu_{\mathrm{disgust,\,sadness}}$   & 0.00 & 0.01 & 0.00 & 0.00\\
$\mu_0^{\mathrm{disgust}}$           & 0.04 & 0.05 & 0.04 & 0.04\\
\midrule
\multicolumn{5}{l}{\textbf{\emph{Fear}}}\\
$\nu_{\mathrm{fear,\,joy}}$        & 0.00 & 0.00 & 0.00 & 0.00\\
$\nu_{\mathrm{fear,\,surprise}}$   & 0.01 & 0.02 & 0.01 & 0.01\\
$\nu_{\mathrm{fear,\,anger}}$      & 0.00 & 0.00 & 0.00 & 0.00\\
$\nu_{\mathrm{fear,\,disgust}}$    & 0.00 & 0.00 & 0.00 & 0.00\\
$\nu_{\mathrm{fear,\,fear}}$       & 0.03 & 0.03 & 0.02 & 0.05\\
$\nu_{\mathrm{fear,\,sadness}}$    & 0.01 & 0.02 & 0.01 & 0.01\\
$\mu_0^{\mathrm{fear}}$            & 0.01 & 0.01 & 0.01 & 0.01\\
\midrule
\multicolumn{5}{l}{\textbf{\emph{Sadness}}}\\
$\nu_{\mathrm{sadness,\,joy}}$       & 0.00 & 0.00 & 0.00 & 0.00\\
$\nu_{\mathrm{sadness,\,surprise}}$  & 0.00 & 0.01 & 0.00 & 0.00\\
$\nu_{\mathrm{sadness,\,anger}}$     & 0.00 & 0.01 & 0.00 & 0.00\\
$\nu_{\mathrm{sadness,\,disgust}}$   & 0.00 & 0.00 & 0.00 & 0.00\\
$\nu_{\mathrm{sadness,\,fear}}$      & 0.01 & 0.03 & 0.00 & 0.01\\
$\nu_{\mathrm{sadness,\,sadness}}$   & 0.02 & 0.03 & 0.02 & 0.03\\
$\mu_0^{\mathrm{sadness}}$           & 0.03 & 0.03 & 0.03 & 0.02\\
\bottomrule
\end{tabular}
\end{adjustbox}
\end{minipage}
\end{table*}
\begin{table*}[!htb]
\centering
\caption{
Statistics of emotion dynamics for different model specifications. 
The spectral radius $\rho$ of the branching ratio matrix $\alpha^{e,f}$ provides a global measure of self-excitation and cross-excitation.
For each video, we calculate the average ratio of endogenous (exogenous) intensity to the total intensity across time, and obtain a fraction of endogenous versus exogenous intensity.
For each emotion, we calculate and report the median endogenous vs. exogenous intensity ratio from the distribution of these ratios across videos. 
We obtain the median statistics of spontaneous vs. video-influenced intensity ratios for each emotion in the same manner.
The \textit{Baseline} model corresponds to our main analysis, with results also reported in Figure 3 in the main text.  
For all other models, we report the results using subsets of videos filtered by the specified keywords. 
The relative influences are highly consistent with our main result across parametrizations of video influence. 
The endogenous influence is comparably larger than the exogenous influence, and the source of mother events is predominantly spontaneous user expressions.
}
\label{tab:ratios_vid}
\large
\setlength{\tabcolsep}{6pt}
\renewcommand{\arraystretch}{1.3}
\begin{tabular}{lcccc}
\toprule
\textbf{Statistic} & \textbf{Baseline} & \textbf{Factor 0.5} & \textbf{Factor 2} & \textbf{Power Law} \\
\midrule
\textbf{\textit{Spectral radius}} & 0.80 & 0.79 & 0.79 & 0.80 \\
\midrule
\multicolumn{5}{l}{\textbf{\textit{Endogenous vs.\ Exogenous Intensity}}} \\
\quad joy       & 3.05 & 3.06 & 3.00 & 3.10 \\
\quad surprise  & 1.07 & 1.07 & 1.01 & 1.05 \\
\quad anger     & 1.80 & 1.87 & 1.79 & 1.73 \\
\quad disgust   & 1.93 & 2.01 & 1.94 & 1.88 \\
\quad fear      & 1.77 & 1.73 & 1.63 & 1.79 \\
\quad sadness   & 2.71 & 2.67 & 2.62 & 2.62 \\
\midrule
\multicolumn{5}{l}{\textbf{\textit{Spontaneous Expression vs.\ Video-Induced}}} \\
\quad joy       & 4.08 & 4.64 & 3.91 & 8.74 \\
\quad surprise  & 2.41 & 3.16 & 3.01 & 2.60 \\
\quad anger     & 4.17 & 6.09 & 2.59 & 2.35 \\
\quad disgust   & 4.51 & 6.69 & 2.81 & 2.68 \\
\quad fear      & 2.26 & 3.36 & 1.53 & 0.93 \\
\quad sadness   & 6.08 & 6.94 & 5.63 & 4.28 \\
\bottomrule
\end{tabular}
\end{table*}

To examine the impact of video content on live chat discussions, we have parameterized and interpolated the prevalence of emotions in the video via sums of log-normal shapes (see Methods). 
Here, we test the robustness of our results with respect to changes in the log-normal parameters. 
We have defined the shape of the log-normal function by calibrating its mode (maximum value), and the median.
We assumed that the intensity of each transcript peaks $2$ seconds after appearance such that the maximum of the log-normal function lies at $2$ seconds. 
We further assumed that 50\% of the emotion intensity for each transcript is manifested within $10$ seconds of transcript appearance such that the median is equal to $10$ seconds.
Here we repeat our analysis while changing the log-normal function maximum and median by factors of 0.5 and 2 to test the robustness of our results.
This effectively changes assumptions about the audience's reaction time and memory of prior content.
We report the results of these analyses in columns Factor 0.5 and Factor 2 of Supplementary Tables \ref{tab:param_vid} and \ref{tab:ratios_vid}.
We observe that the results are qualitatively consistent with the results in the main text.
This suggests that our results aren't particularly sensitive to the parametrization of the video content. 

Human memory decays following an approximate power law \cite{pollmann1998forgetting}.
The log-normal function resembles the shape of a power law over a limited range of the variable, which is all the larger, the larger is $\sigma$ \cite{sornette2006critical}.
We thus test our results with an alternative parametrization of the video influence in a power law shape where $s^{f}_{\tau_j}(t) = \frac{1}{(t - \tau_j)^c}$.
We report results in column Power Law of Supplementary Tables \ref{tab:param_vid} and \ref{tab:ratios_vid} for $c = 2.5$, and find that the results are qualitatively consistent with the results we obtain from the log-normal over a range of $c$ values from $2.5$ to $10$.

\subsection{Robustness Check on Livechat Filtering} 
\label{sec:robust_livechat}

\begin{table*}[!htb]
\centering
\large
\setlength{\tabcolsep}{4.5pt}
\renewcommand{\arraystretch}{1.12}
\caption{
Mean estimates of endogenous excitation parameters $\alpha_{e,f}$ and decay rates $\gamma^{e}$ (left), and exogenous influence parameters $\nu_{e,f}$ and spontaneous baselines $\mu_0^{e}$ (right) for different model specfications. 
The \textit{Baseline} model corresponds to our main analysis, with results also reported in Figure 2 in the main text.  
We conduct several robustness checks by varying key preprocessing steps and model configurations.  
We further conduct robustness checks using alternative data processing and modeling choices.  
In \textbf{Model A}, we include the full sample of live chat messages without filtering out messages sent before or after the video duration.  
In \textbf{Model B}, we exclude videos that are manually identified as centering around direct interactions between the live streamer and the audience.
In the \textbf{4 Emotions} specification, we restrict the analysis to a reduced set of emotions ($|\mathcal{E}|=4$), modeling only \textit{joy}, \textit{anger}, \textit{disgust}, and \textit{sadness}, while excluding \textit{fear} and \textit{surprise}.  
These alternative specifications help validate the robustness of our core findings. 
}
\label{tab:param_add}
\begin{minipage}[t]{0.49\textwidth}
\centering
\subcaption*{\textbf{(a) Endogenous $\alpha_{e,f}$ and $\gamma^{e}$}}
\begin{adjustbox}{max width=\linewidth}
\begin{tabular}{lcccc}
\toprule
\textbf{Statistic} & \textbf{Baseline} & \textbf{Model A} & \textbf{Model B} & \textbf{4 Emotions} \\
\midrule
\multicolumn{5}{l}{\textbf{\emph{Joy}}}\\
$\alpha_{\mathrm{joy,joy}}$            & 0.76 & 0.83 & 0.73 & 0.84\\
$\alpha_{\mathrm{joy,surprise}}$       & 0.03 & 0.05 & 0.03 & /\\
$\alpha_{\mathrm{joy,anger}}$          & 0.00 & 0.01 & 0.07 & 0.00\\
$\alpha_{\mathrm{joy,disgust}}$        & 0.20 & 0.28 & 0.21 & 0.23\\
$\alpha_{\mathrm{joy,fear}}$           & 0.08 & 0.15 & 0.13 & /\\
$\alpha_{\mathrm{joy,sadness}}$        & 0.21 & 0.24 & 0.18 & 0.22\\
$\gamma^{\text{joy}}$                  & 2.36 & 2.68 & 2.28 & 2.39\\
\midrule
\multicolumn{5}{l}{\textbf{\emph{Surprise}}}\\
$\alpha_{\mathrm{surprise,joy}}$       & 0.01 & 0.01 & 0.00 & /\\
$\alpha_{\mathrm{surprise,surprise}}$  & 0.05 & 0.05 & 0.06 & /\\
$\alpha_{\mathrm{surprise,anger}}$     & 0.01 & 0.02 & 0.00 & /\\
$\alpha_{\mathrm{surprise,disgust}}$   & 0.01 & 0.01 & 0.01 & /\\
$\alpha_{\mathrm{surprise,fear}}$      & 0.01 & 0.01 & 0.02 & /\\
$\alpha_{\mathrm{surprise,sadness}}$   & 0.01 & 0.01 & 0.00 & /\\
$\gamma^{\text{surprise}}$             & 3.44 & 4.90 & 2.07 & /\\
\midrule
\multicolumn{5}{l}{\textbf{\emph{Anger}}}\\
$\alpha_{\mathrm{anger,joy}}$          & 0.02 & 0.02 & 0.03 & 0.02\\
$\alpha_{\mathrm{anger,surprise}}$     & 0.07 & 0.08 & 0.10 & /\\
$\alpha_{\mathrm{anger,anger}}$        & 0.26 & 0.27 & 0.28 & 0.36\\
$\alpha_{\mathrm{anger,disgust}}$      & 0.20 & 0.21 & 0.14 & 0.26\\
$\alpha_{\mathrm{anger,fear}}$         & 0.10 & 0.11 & 0.11 & /\\
$\alpha_{\mathrm{anger,sadness}}$      & 0.04 & 0.05 & 0.06 & 0.06\\
$\gamma^{\text{anger}}$                & 4.81 & 5.25 & 4.69 & 5.31\\
\midrule
\multicolumn{5}{l}{\textbf{\emph{Disgust}}}\\
$\alpha_{\mathrm{disgust,joy}}$        & 0.02 & 0.03 & 0.03 & 0.02\\
$\alpha_{\mathrm{disgust,surprise}}$   & 0.05 & 0.09 & 0.10 & /\\
$\alpha_{\mathrm{disgust,anger}}$      & 0.20 & 0.21 & 0.20 & 0.27\\
$\alpha_{\mathrm{disgust,disgust}}$    & 0.26 & 0.27 & 0.21 & 0.34\\
$\alpha_{\mathrm{disgust,fear}}$       & 0.09 & 0.13 & 0.11 & /\\
$\alpha_{\mathrm{disgust,sadness}}$    & 0.06 & 0.08 & 0.09 & 0.09\\
$\gamma^{\text{disgust}}$              & 4.64 & 5.24 & 4.63 & 5.23\\
\midrule
\multicolumn{5}{l}{\textbf{\emph{Fear}}}\\
$\alpha_{\mathrm{fear,joy}}$           & 0.01 & 0.01 & 0.01 & /\\
$\alpha_{\mathrm{fear,surprise}}$      & 0.02 & 0.04 & 0.03 & /\\
$\alpha_{\mathrm{fear,anger}}$         & 0.02 & 0.02 & 0.02 & /\\
$\alpha_{\mathrm{fear,disgust}}$       & 0.00 & 0.00 & 0.00 & /\\
$\alpha_{\mathrm{fear,fear}}$          & 0.12 & 0.14 & 0.12 & /\\
$\alpha_{\mathrm{fear,sadness}}$       & 0.04 & 0.04 & 0.04 & /\\
$\gamma^{\text{fear}}$                 & 3.33 & 4.27 & 3.14 & /\\
\midrule
\multicolumn{5}{l}{\textbf{\emph{Sadness}}}\\
$\alpha_{\mathrm{sadness,joy}}$        & 0.03 & 0.03 & 0.03 & 0.03\\
$\alpha_{\mathrm{sadness,surprise}}$   & 0.06 & 0.06 & 0.05 & /\\
$\alpha_{\mathrm{sadness,anger}}$      & 0.01 & 0.01 & 0.02 & 0.01\\
$\alpha_{\mathrm{sadness,disgust}}$    & 0.07 & 0.10 & 0.07 & 0.11\\
$\alpha_{\mathrm{sadness,fear}}$       & 0.11 & 0.13 & 0.12 & /\\
$\alpha_{\mathrm{sadness,sadness}}$    & 0.21 & 0.24 & 0.19 & 0.26\\
$\gamma^{\text{sadness}}$              & 3.61 & 4.02 & 3.42 & 3.60\\
\bottomrule
\end{tabular}
\end{adjustbox}
\end{minipage}
%
\hfill
\begin{minipage}[t]{0.49\textwidth}
\centering
\subcaption*{\textbf{(b) Exogenous $\nu_{e,f}$ and $\mu_{0}^{e}$}}
\begin{adjustbox}{max width=\linewidth}
\begin{tabular}{lcccc}
\toprule
\textbf{Statistic} & \textbf{Baseline} & \textbf{Model A} & \textbf{Model B} & \textbf{4 Emotions} \\
\midrule
\multicolumn{5}{l}{\textbf{\emph{Joy}}}\\
$\nu_{\mathrm{joy,joy}}$            & 0.03 & 0.04 & 0.04 & 0.02\\
$\nu_{\mathrm{joy,surprise}}$       & 0.05 & 0.07 & 0.02 & /\\
$\nu_{\mathrm{joy,anger}}$          & 0.00 & 0.00 & 0.00 & 0.00\\
$\nu_{\mathrm{joy,disgust}}$        & 0.00 & 0.00 & 0.00 & 0.00\\
$\nu_{\mathrm{joy,fear}}$           & 0.00 & 0.01 & 0.01 & /\\
$\nu_{\mathrm{joy,sadness}}$        & 0.00 & 0.00 & 0.00 & 0.00\\
$\mu_0^{\text{joy}}$                & 0.29 & 0.15 & 0.26 & 0.20\\
\midrule
\multicolumn{5}{l}{\textbf{\emph{Surprise}}}\\
$\nu_{\mathrm{surprise,joy}}$       & 0.00 & 0.00 & 0.00 & /\\
$\nu_{\mathrm{surprise,surprise}}$  & 0.00 & 0.00 & 0.00 & /\\
$\nu_{\mathrm{surprise,anger}}$     & 0.00 & 0.00 & 0.00 & /\\
$\nu_{\mathrm{surprise,disgust}}$   & 0.00 & 0.00 & 0.00 & /\\
$\nu_{\mathrm{surprise,fear}}$      & 0.00 & 0.00 & 0.01 & /\\
$\nu_{\mathrm{surprise,sadness}}$   & 0.00 & 0.00 & 0.01 & /\\
$\mu_0^{\text{surprise}}$           & 0.01 & 0.00 & 0.01 & /\\
\midrule
\multicolumn{5}{l}{\textbf{\emph{Anger}}}\\
$\nu_{\mathrm{anger,joy}}$          & 0.00 & 0.00 & 0.00 & 0.00\\
$\nu_{\mathrm{anger,surprise}}$     & 0.02 & 0.04 & 0.02 & /\\
$\nu_{\mathrm{anger,anger}}$        & 0.03 & 0.03 & 0.02 & 0.03\\
$\nu_{\mathrm{anger,disgust}}$      & 0.01 & 0.02 & 0.01 & 0.01\\
$\nu_{\mathrm{anger,fear}}$         & 0.01 & 0.01 & 0.01 & /\\
$\nu_{\mathrm{anger,sadness}}$      & 0.00 & 0.00 & 0.00 & 0.00\\
$\mu_0^{\text{anger}}$              & 0.04 & 0.02 & 0.04 & 0.03\\
\midrule
\multicolumn{5}{l}{\textbf{\emph{Disgust}}}\\
$\nu_{\mathrm{disgust,joy}}$        & 0.00 & 0.00 & 0.00 & 0.00\\
$\nu_{\mathrm{disgust,surprise}}$   & 0.03 & 0.04 & 0.04 & /\\
$\nu_{\mathrm{disgust,anger}}$      & 0.02 & 0.02 & 0.02 & 0.02\\
$\nu_{\mathrm{disgust,disgust}}$    & 0.01 & 0.02 & 0.01 & 0.01\\
$\nu_{\mathrm{disgust,fear}}$       & 0.02 & 0.02 & 0.02 & /\\
$\nu_{\mathrm{disgust,sadness}}$    & 0.00 & 0.00 & 0.00 & 0.00\\
$\mu_0^{\text{disgust}}$            & 0.04 & 0.02 & 0.04 & 0.03\\
\midrule
\multicolumn{5}{l}{\textbf{\emph{Fear}}}\\
$\nu_{\mathrm{fear,joy}}$           & 0.00 & 0.00 & 0.00 & /\\
$\nu_{\mathrm{fear,surprise}}$      & 0.01 & 0.01 & 0.00 & /\\
$\nu_{\mathrm{fear,anger}}$         & 0.00 & 0.00 & 0.00 & /\\
$\nu_{\mathrm{fear,disgust}}$       & 0.00 & 0.01 & 0.00 & /\\
$\nu_{\mathrm{fear,fear}}$          & 0.03 & 0.03 & 0.04 & /\\
$\nu_{\mathrm{fear,sadness}}$       & 0.01 & 0.01 & 0.01 & /\\
$\mu_0^{\text{fear}}$               & 0.01 & 0.00 & 0.01 & /\\
\midrule
\multicolumn{5}{l}{\textbf{\emph{Sadness}}}\\
$\nu_{\mathrm{sadness,joy}}$        & 0.00 & 0.00 & 0.00 & 0.00\\
$\nu_{\mathrm{sadness,surprise}}$   & 0.00 & 0.00 & 0.00 & /\\
$\nu_{\mathrm{sadness,anger}}$      & 0.00 & 0.00 & 0.01 & 0.00\\
$\nu_{\mathrm{sadness,disgust}}$    & 0.00 & 0.00 & 0.00 & 0.00\\
$\nu_{\mathrm{sadness,fear}}$       & 0.01 & 0.02 & 0.02 & /\\
$\nu_{\mathrm{sadness,sadness}}$    & 0.02 & 0.02 & 0.01 & 0.02\\
$\mu_0^{\text{sadness}}$            & 0.03 & 0.01 & 0.03 & 0.02\\
\bottomrule
\end{tabular}
\end{adjustbox}
\end{minipage}
\end{table*}
\begin{table*}[!htb]
\centering
\caption{
Statistics of emotion dynamics for different model specifications. 
The spectral radius $\rho$ of the branching ratio matrix $\alpha^{e,f}$ provides a global measure of self-excitation and cross-excitation.
For each video, we calculate the average ratio of endogenous (exogenous) intensity to the total intensity across time, and obtain a fraction of endogenous versus exogenous intensity.
For each emotion, we calculate and report the median endogenous vs. exogenous intensity ratio from the distribution of these ratios across videos. 
We obtain the median statistics of spontaneous vs. video-influenced intensity ratios for each emotion in the same manner.
The \textit{Baseline} model corresponds to our main analysis, with results also reported in Figure 3 in the main text.  
In \textbf{Model A}, we include the full sample of live chat messages without filtering out messages sent before or after the video duration.  
In \textbf{Model B}, we exclude videos that are manually identified as centering around direct interactions between the live streamer and the audience.
In the \textbf{4 Emotions} specification, we restrict the analysis to a reduced set of emotions ($|\mathcal{E}|=4$), modeling only \textit{joy}, \textit{anger}, \textit{disgust}, and \textit{sadness}, while excluding \textit{fear} and \textit{surprise}.  
}
\label{tab:ratios_add}
\large
\setlength{\tabcolsep}{6pt}
\renewcommand{\arraystretch}{1.3}
\begin{tabular}{lcccc}
\toprule
\textbf{Statistic} & \textbf{Baseline} & \textbf{Model A} & \textbf{Model B} & \textbf{4 Emotions} \\
\midrule
\textbf{\textit{Spectral radius}} & 0.80 & 0.87 & 0.77 & 0.88 \\
\midrule
\multicolumn{5}{l}{\textbf{\textit{Endogenous vs.\ Exogenous Intensity}}} \\
\quad joy       & 3.05 & 4.84 & 2.84 & 4.09 \\
\quad surprise  & 1.07 & 2.53 & 0.59 & /    \\
\quad anger     & 1.80 & 3.05 & 2.11 & 2.65 \\
\quad disgust   & 1.93 & 3.33 & 2.15 & 2.83 \\
\quad fear      & 1.77 & 3.38 & 1.74 & /    \\
\quad sadness   & 2.71 & 4.99 & 2.15 & 3.90 \\
\midrule
\multicolumn{5}{l}{\textbf{\textit{Spontaneous Expression vs.\ Video-Induced}}} \\
\quad joy       & 4.08 & 2.20 & 2.64 & 4.15 \\
\quad surprise  & 2.41 & 1.26 & 4.18 & /    \\
\quad anger     & 4.17 & 1.69 & 3.75 & 3.86 \\
\quad disgust   & 4.51 & 1.48 & 3.57 & 4.30 \\
\quad fear      & 2.26 & 0.93 & 1.96 & /    \\
\quad sadness   & 6.08 & 2.47 & 5.18 & 5.81 \\
\bottomrule
\end{tabular}
\end{table*}

In the main analysis, we restrict the observation period of the analysis to the time window covered by the video transcript, excluding the pre- and post-transcript periods, for the following reasons.
A central advantage of using YouTube Live videos in our setting is the presence of an observable exogenous influence from the emotional content of the video itself. 
Once the video ends, however, this exogenous signal is no longer reliably observable, and viewers are more likely to be influenced by other external factors that are beyond the scope of our model.
Additionally, live chat activity typically drops off sharply after the video concludes, often becoming sparse or irregular, which could introduce additional noise and make parameter estimation less reliable.

Nonetheless, we examine the robustness of our findings with respect to the filtering of live chat messages.
We use an updated data sample without filtering out live chat messages that occurred before or after the official video runtime for each video. 
The resulting observation windows for each video span from the timestamp of the first live chat message to that of the last. 
The video influence at each time point is computed following the same procedure detailed in the Methods section. 
The results of this extended analysis are presented in column Model A of Supplementary Tables \ref{tab:param_add} and \ref{tab:ratios_add}.
We find that including the full set of live chats does not alter our main conclusions, further reinforcing the robustness of our findings.

\subsection{Robustness Check on Live Interactions} 
\label{sec:robust_live}

In our main analysis, we assume that the direction of influence in live videos flows predominantly from the video content to the audience.  
However, livestreams can vary significantly in format, ranging from passive broadcasts to highly interactive sessions that involve real-time exchanges with viewers.  
The validity of our unidirectional influence assumption may therefore depend on the type of livestream. 
In the following, we examine whether different types of livestream formats may affect our findings.  
To do so, we manually labeled all 397 videos in our sample into one of three categories: 

\begin{itemize}
  \item[\(2\)] \textbf{Interactive live streams:} Videos where the content actively involves and often centers around direct interactions between the live streamer and the audience. For example, a live Q\&A session such as \href{https://www.youtube.com/watch?v=2xCfJBckRTQ}{this video}.
  
  \item[\(1\)] \textbf{Live streams with potential, but limited, interaction:} Videos in which audience interaction could theoretically occur, but is not a core part of the video’s format or purpose. For example, a live commentary show such as \href{https://www.youtube.com/watch?v=4RAk1TdBLz8}{this video}.
  
  \item[\(0\)] \textbf{Pre-recorded or passive streams:} Videos in which the live audience cannot influence the video content. For example, a live sports competition such as \href{https://www.youtube.com/watch?v=-8rcsRjEvCU}{this video}.
\end{itemize}

Based on our manual labeling, approximately 9\% of the videos in our sample are either pre-recorded or passive streams where audience interaction is not possible (category 0). 
About 61\% of the videos fall into category 1, where audience interaction is theoretically possible but not central to the video’s intent. 
The remaining 30\% of the videos are in category 2, where real-time interaction between the streamer and the audience is likely and the assumption of unidirectional influence is most questionable. 

To examine the sensitivity of our results to this assumption, we excluded all fully interactive videos (label = 2) and re-estimated the model on the remaining 241 videos.  
The results from this restricted sample are reported under ``Model B'' in Supplementary Tables~\ref{tab:param_add} and~\ref{tab:ratios_add}.  
We find that the estimates remain qualitatively consistent with our main findings.  
\coloredtext{
Moreover, the numerical differences between the baseline model and Model B generally fall well within one standard deviation of our bootstrapped parameter estimates, suggesting that the observed variation is likely due to statistical noise.  
To formally verify this, we conducted two-sample t-tests comparing the parameter distributions of the baseline and restricted models across all six emotions and for all parameters, $\mu^{e}_0$, $\nu^{e,f}$, $\alpha^{e,f}$, and $\gamma^e$.  
For all cases, the resulting $p$-values were greater than $0.1$, indicating that we cannot reject the null hypothesis that the two sets of parameter estimates are drawn from the same underlying distribution.  
These results confirm that the presence or absence of fully interactive videos does not meaningfully affect our overall conclusions.
}

Additionally, we note that if the audience does influence the live streamer, then the video content itself becomes partially endogenous to the system. 
This introduces a circular dynamic: chat influences video, which in turn influences chat. 
In such cases, the video - originally treated as an exogenous source  - would, in fact, contain endogenous components. 
This feedback loop would further strengthen our core conclusion: endogenous dynamics, whether direct (chat-to-chat) or mediated via the video, play a dominant role in shaping emotional expression during live interactions.

\subsection{Dynamics of 4 Emotions} 
\label{sec:robust_4}

Our main analysis models the commonly studied 6 basic emotions \cite{ekman1992}. 
Here we test the robustness of our results by modeling $|\mathcal{E}|=4$ emotions: joy, anger, disgust, and sadness, removing \textit{fear} and \textit{surprise} 
because they have small excitatory patterns and low overall intensity.
We obtain a sample of 288,581 live chat messages across 1,405 videos.
Reducing the dimensionality of the multivariate Hawkes process reduces the number of parameters to be estimated from 84 to 40.
We report our results for 4 emotions in column 4 Emotions of Supplementary Tables \ref{tab:param_add} and \ref{tab:ratios_add}.
We find that the dynamics of 4 emotions are qualitatively similar to our main results on 6 emotions (Figures 2 and 3 in the main text).

\subsection{Dynamics of Live Chats with No Emotional Content} 
\label{sec:no_emotion}

\begin{table*}[!htb]
    \centering
    \begin{tabular}{|c|c|c|c|c|}
        \hline
        Parameters:            & $\alpha$    &  $\gamma$ &  $\mu$   & $\nu$ \\
        \hline
        Mean                       & 0.857         & 1.967          & 0.150     & 0.029 \\
        \hline
        Standard Deviation & 0.018          & 0.167          & 0.022     & 0.006 \\
        \hline
    \end{tabular}
    \caption{The mean and standard deviation of estimated parameters for a univariate Hawkes Process of live chat messages with no emotion labels across 10 bootstrapped samples.}
    \label{tab:no_emotion_param}
\end{table*}

\begin{figure*}[!htb]
	\centering
	\includegraphics[width=\textwidth]{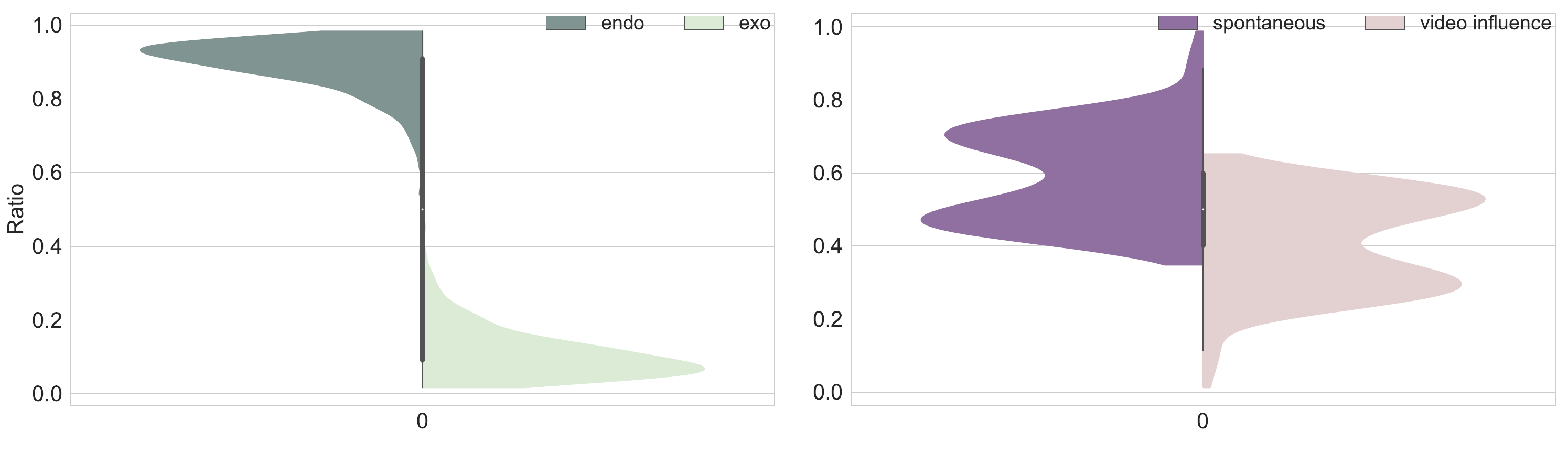}
	\caption{
			The left plot shows the relative ratios of endogenous and exogenous influences for live chat messages with no emotion labels.
			The right plot shows the ratios of mother events from spontaneous expression and video influence for live chat messages with no emotion labels.
			We observe that the endogenous influence nonetheless dominates the dynamics of live chat messages with no emotion labels.
			However, the source of mother events is distributed more evenly between spontaneous expressions and video influence.
		}
	\label{fig:ratio_no_emotion}
\end{figure*}

Our main study shows that online emotions are characterized by excitatory patterns.
Despite the constant feed of video content, the emotion dynamics evolve mostly spontaneously among users.
To complement our analysis, we study the behavior of live chats without emotional content.
We train a transformer to assign emotion labels for live chats.
We consider live chat messages that were not assigned any emotion labels to contain no emotional content.
This gives us 66,235 live chat messages with no emotional content across 232 videos.
We use a univariate Hawkes process to describe the intensity of live chat messages without emotion labels.
The intensity $\lambda(t)$ at time $t$ during a given video is specified as
\begin{equation}
\scriptstyle 
	\lambda(t)
	= 
	\underbrace{
	\mu_0
	}_{\text{exo base rate}}
	+ 
	\underbrace{
	\nu S(t)
	}_{\text{exo video influence}}
	+ 
	\underbrace{
	\sum_{t_j < t} \phi(t-t_j)
	}_{\text{endo chat influence}}
	\label{eq:uni_hawkes_no_emotions}.
\end{equation} 
The value of $S(t)$ is given by $S(t) = \sum_{\tau_j < t} s_{\tau_j}(t)$ where $\left\{ \tau_j \right\}$ enumerates all times at which a subtitle first appears in the video.
The values of $s_{\tau_j}(t)$ are calculated according to Equation (3) from the main text with $\mu= 2.3$ and $\sigma = 1.3$.
The term $s_{\tau_j}(t)$ represents the time-varying intensity arising from a video subscript appearing in the video at time $\tau_j$. 
Similarly, we assume that $\phi$ is exponentially decaying where $\phi(t) = \alpha \frac{1}{\gamma} e^{-\frac{1}{\gamma} t}$ with decay time $\gamma$.
Supplementary Table \ref{tab:no_emotion_param} shows the estimated parameters of Equation \eqref{eq:uni_hawkes_no_emotions}.
We observe that live chat messages with no emotional content are nonetheless highly self-exciting but with a shorter direct memory than emotional live chats.
Unlike our main result (Figure 3 in the main text), live chat messages with no emotional content are driven more equally by the video and user spontaneity, as shown in Supplementary Figure \ref{fig:ratio_no_emotion}.
This shows that while live discussions evolve largely endogenously, live chats with no emotional content are much more affected by videos.
In contrast, emotions on online platforms are largely attributed to user expressions and interactions.

\end{appendices}


\end{document}